\documentclass[twocolumn]{aastex631}

\usepackage{amsmath}

\usepackage{natbib}
\defcitealias{Zawadzki_2023}{Z23}

\newcommand{\disksurf}{\texttt{disksurf}}

\shorttitle{exoALMA: RML}
\shortauthors{Zawadzki et al.}

\begin{document}

\title{exoALMA IX: Regularized Maximum Likelihood Imaging of Non-Keplerian Features}


\author[0000-0001-9319-1296]{Brianna Zawadzki}
\affiliation{Department of Astronomy, Van Vleck Observatory, Wesleyan University, 96 Foss Hill Drive, Middletown, CT 06459, USA}
\affiliation{Department of Astronomy \& Astrophysics, 525 Davey Laboratory, The Pennsylvania State University, University Park, PA 16802, USA}

\author[0000-0002-1483-8811]{Ian Czekala}
\affiliation{School of Physics \& Astronomy, University of St. Andrews, North Haugh, St. Andrews KY16 9SS, UK}
\affiliation{Centre for Exoplanet Science, University of St. Andrews, North Haugh, St. Andrews, KY16 9SS, UK}

\author[0000-0002-5503-5476]{Maria Galloway-Sprietsma}
\affiliation{Department of Astronomy, University of Florida, Gainesville, FL 32611, USA}

\author[0000-0001-7258-770X]{Jaehan Bae}
\affiliation{Department of Astronomy, University of Florida, Gainesville, FL 32611, USA}

\author[0000-0001-6378-7873]{Marcelo Barraza-Alfaro}
\affiliation{Department of Earth, Atmospheric, and Planetary Sciences, Massachusetts Institute of Technology, Cambridge, MA 02139, USA}

\author[0000-0002-7695-7605]{Myriam Benisty}
\affiliation{Max-Planck Institute for Astronomy (MPIA), Königstuhl 17, D-69117 Heidelberg, Germany}
\affiliation{Univ. Grenoble Alpes, CNRS, IPAG, 38000 Grenoble, France}
\affiliation{Universit\'e C\^ote d'Azur, Observatoire de la C\^ote d'Azur, CNRS, Laboratoire Lagrange, France}

\author[0000-0002-2700-9676]{Gianni Cataldi} 
\affiliation{National Astronomical Observatory of Japan, Osawa 2-21-1, Mitaka, Tokyo 181-8588, Japan}

\author[0000-0003-2045-2154]{Pietro Curone} 
\affiliation{Dipartimento di Fisica, Universit\`a degli Studi di Milano, Via Celoria 16, 20133 Milano, Italy}
\affiliation{Departamento de Astronom\'ia, Universidad de Chile, Camino El Observatorio 1515, Las Condes, Santiago, Chile}

\author[0000-0003-4689-2684]{Stefano Facchini}
\affiliation{Dipartimento di Fisica, Universit\`a degli Studi di Milano, Via Celoria 16, 20133 Milano, Italy}

\author[0000-0003-4679-4072]{Daniele Fasano} 
\affiliation{Univ. Grenoble Alpes, CNRS, IPAG, 38000 Grenoble, France}
\affiliation{Universit\'e C\^ote d'Azur, Observatoire de la C\^ote d'Azur, CNRS, Laboratoire Lagrange, France}

\author[0000-0002-9298-3029]{Mario Flock} 
\affiliation{Max-Planck Institute for Astronomy (MPIA), Königstuhl 17, D-69117 Heidelberg, Germany}

\author[0000-0003-1117-9213]{Misato Fukagawa} 
\affiliation{National Astronomical Observatory of Japan, Osawa 2-21-1, Mitaka, Tokyo 181-8588, Japan}

\author[0000-0002-5910-4598]{Himanshi Garg}
\affiliation{School of Physics and Astronomy, Monash University, Clayton VIC 3800, Australia}

\author[0000-0002-8138-0425]{Cassandra Hall}
\affiliation{1 Department of Physics and Astronomy, The University of Georgia, Athens, GA 30602, USA}
\affiliation{Center for Simulational Physics, The University of Georgia, Athens, GA 30602, USA}
\affiliation{Institute for Artificial Intelligence, The University of Georgia, Athens, GA, 30602, USA}

\author[0000-0001-7641-5235]{Thomas Hilder}
\affiliation{School of Physics and Astronomy, Monash University, VIC 3800, Australia}

\author[0000-0001-6947-6072]{Jane Huang} 
\affiliation{Department of Astronomy, Columbia University, 538 W. 120th Street, Pupin Hall, New York, NY, USA}

\author[0000-0003-1008-1142]{John~D.~Ilee} 
\affiliation{School of Physics and Astronomy, University of Leeds, Leeds, UK, LS2 9JT}

\author[0000-0001-8061-2207]{Andrea Isella}
\affiliation{Department of Physics and Astronomy, Rice University, 6100 Main St, Houston, TX 77005, USA}
\affiliation{Rice Space Institute, Rice University, 6100 Main St, Houston, TX 77005, USA}

\author[0000-0001-8446-3026]{Andr\'es F. Izquierdo} 
\affiliation{Department of Astronomy, University of Florida, Gainesville, FL 32611, USA}
\affiliation{Leiden Observatory, Leiden University, P.O. Box 9513, NL-2300 RA Leiden, The Netherlands}
\affiliation{European Southern Observatory, Karl-Schwarzschild-Str. 2, D-85748 Garching bei M\"unchen, Germany}
\affiliation{NASA Hubble Fellowship Program Sagan Fellow}

\author[0000-0001-7235-2417]{Kazuhiro Kanagawa} 
\affiliation{College of Science, Ibaraki University, 2-1-1 Bunkyo, Mito, Ibaraki 310-8512, Japan}

\author[0000-0002-8896-9435]{Geoffroy Lesur} 
\affiliation{Univ. Grenoble Alpes, CNRS, IPAG, 38000, Grenoble, France}

\author[0000-0003-4663-0318]{Cristiano Longarini} 
\affiliation{Institute of Astronomy, University of Cambridge, Madingley Road, CB3 0HA, Cambridge, UK}
\affiliation{Dipartimento di Fisica, Universit\`a degli Studi di Milano, Via Celoria 16, 20133 Milano, Italy}

\author[0000-0002-8932-1219]{Ryan A. Loomis}
\affiliation{National Radio Astronomy Observatory, 520 Edgemont Rd., Charlottesville, VA 22903, USA}


\author[0000-0003-4039-8933]{Ryuta Orihara} 
\affiliation{College of Science, Ibaraki University, 2-1-1 Bunkyo, Mito, Ibaraki 310-8512, Japan}

\author[0000-0001-5907-5179]{Christophe Pinte}
\affiliation{School of Physics and Astronomy, Monash University, Clayton VIC 3800, Australia}
\affiliation{Univ. Grenoble Alpes, CNRS, IPAG, 38000 Grenoble, France}

\author[0000-0002-4716-4235]{Daniel J. Price} 
\affiliation{School of Physics and Astronomy, Monash University, Clayton VIC 3800, Australia}

\author[0000-0003-4853-5736]{Giovanni Rosotti} 
\affiliation{Dipartimento di Fisica, Universit\`a degli Studi di Milano, Via Celoria 16, 20133 Milano, Italy}

\author[0000-0002-0491-143X]{Jochen Stadler} 
\affiliation{Universit\'e C\^ote d'Azur, Observatoire de la C\^ote d'Azur, CNRS, Laboratoire Lagrange, 06304 Nice, France}
\affiliation{Univ. Grenoble Alpes, CNRS, IPAG, 38000 Grenoble, France}

\author[0000-0003-1534-5186]{Richard Teague}
\affiliation{Department of Earth, Atmospheric, and Planetary Sciences, Massachusetts Institute of Technology, Cambridge, MA 02139, USA}

\author[0000-0003-1412-893X]{Hsi-Wei Yen} 
\affiliation{Academia Sinica Institute of Astronomy \& Astrophysics, 11F of Astronomy-Mathematics Building, AS/NTU, No.1, Sec. 4, Roosevelt Rd, Taipei 10617, Taiwan}

\author[0000-0002-3468-9577]{Gaylor Wafflard-Fernandez} 
\affiliation{Univ. Grenoble Alpes, CNRS, IPAG, 38000 Grenoble, France}

\author[0000-0003-1526-7587]{David J. Wilner}
\affiliation{Center for Astrophysics | Harvard \& Smithsonian, Cambridge, MA 02138, USA}

\author[0000-0002-7501-9801]{Andrew J. Winter}
\affiliation{Universit\'{e} C\^{o}te d'Azur, Observatoire de la C\^{o}te d'Azur, CNRS, Laboratoire Lagrange, 06300 Nice, France}

\author[0000-0002-7212-2416]{Lisa W\"olfer} 
\affiliation{Department of Earth, Atmospheric, and Planetary Sciences, Massachusetts Institute of Technology, Cambridge, MA 02139, USA}

\author[0000-0001-8002-8473]{Tomohiro C. Yoshida}
\affiliation{National Astronomical Observatory of Japan, 2-21-1 Osawa, Mitaka, Tokyo 181-8588, Japan}
\affiliation{Department of Astronomical Science, The Graduate University for Advanced Studies, SOKENDAI, 2-21-1 Osawa, Mitaka, Tokyo 181-8588, Japan}

\begin{abstract}
The planet-hunting ALMA large program exoALMA observed 15 protoplanetary disks at $\sim0\farcs15$ angular resolution and $\sim100$ m/s spectral resolution, characterizing disk structures and kinematics in enough detail to detect non-Keplerian features (NKFs) in the gas emission. As these features are often small and low-contrast, robust imaging procedures are critical for identifying and characterizing NKFs, including determining which features may be signatures of young planets. The exoALMA collaboration employed two different imaging procedures to ensure the consistent detection of NKFs: CLEAN, the standard iterative deconvolution algorithm, and regularized maximum likelihood (RML) imaging. This paper presents the exoALMA RML images, obtained by maximizing the likelihood of the visibility data given a model image and subject to regularizer penalties. Crucially, in the context of exoALMA, RML images serve as an independent verification of marginal features seen in the fiducial CLEAN images. However, best practices for synthesizing RML images of multi-channeled (i.e. velocity-resolved) data remain undefined, as prior work on RML imaging for protoplanetary disk data has primarily addressed single-image cases. We used the open source Python package \texttt{MPoL} to explore RML image validation methods for multi-channeled data and synthesize RML images from the exoALMA observations of 7 protoplanetary disks with apparent NKFs in the $^{12}$CO J=3-2 CLEAN images. We find that RML imaging methods independently reproduce the NKFs seen in the CLEAN images of these sources, suggesting that the NKFs are robust features rather than artifacts from a specific imaging procedure.
\end{abstract}


\section{Introduction}

The Cycle 8 ALMA large program exoALMA has observed 15 protoplanetary disks at sufficient angular and spectral resolution to search for kinematic signatures of young planets by identifying localized deviations from Keplerian velocities, or non-Keplerian features (NKFs), in the disk emission \citep{Teague_exoALMA}. A key objective of the exoALMA program is to make reliable detections of NKFs. Interferometers like ALMA are composed of many individual antennas and deliver data in the spatial frequency domain rather than the image domain. Reconstructing images from interferometric data is an ill-posed mathematical problem that requires assumptions to be made about the unsampled spatial frequencies. Thus, recent recommendations for kinematically detecting planets strongly suggest using multiple image synthesis techniques to image the data \citep{DiskDynamicsCollab_2020}. 

The standard ALMA pipeline uses the Common Astronomy Software Applications (CASA) package for data calibration and image synthesis. The CLEAN algorithm (especially the CASA implementation, \texttt{tclean}), is the most common method for imaging radio interferometric data and, in the absence of noise, is equivalent to a least squares fit to the visibilities \citep{Hoegbom_1974,Schwarz_1978,McMullin_2007,Rau_2011, CASA_2022}. CLEAN is a procedural algorithm which builds a model image by identifying the brightest pixels in the dirty image and placing a CLEAN component at the same location in the model image. Then, the CLEAN component is convolved with the dirty beam (the instrument point spread function) and subtracted from the dirty image. At the end of the CLEANing process (e.g.\ after a specified number of iterations or noise threshold is reached) there are two products: the residual image and the model image. The residual image is what remains after all the dirty-beam-convolved CLEAN components have been subtracted out of the original dirty image. The model image, composed of the many CLEAN components, is typically added to the residual image and convolved with a restoring beam (the ``CLEAN beam'') to obtain a final cleaned image. The fiducial exoALMA images were made with CLEAN, as described in \citet{Teague_exoALMA, Loomis_exoALMA}.

However, despite the fact that the ALMA pipeline and users rely primarily on CLEAN for imaging, it is ultimately a heuristic, user-dependent algorithm which attempts to build well-resolved and often complex structures from a set of point source or Gaussian components. While this typically works well enough, the hierarchical ringed substructures commonly seen in disk observations \citep[e.g.][]{Andrews_2018} are particularly poorly matched to the basis sets of points and Gaussians. Thus, the community should consider adopting imaging methods which do not assume resolved sources like disks are represented by a collection of discrete points. One alternative to CLEAN is regularized maximum likelihood (RML) imaging, a forward modeling approach to imaging which aims to identify the most likely set of image pixels given the data and chosen regularizers \citep[e.g.][]{Cornwell_1985, Narayan_1986, Carcamo_2018, EHTCollab_2019}. RML imaging is an optimization-based imaging framework; the image is obtained by maximizing a likelihood function subject to regularizer penalties. The precise form of the likelihood function can vary depending on the data characteristics (e.g.\ noise qualities) and which regularizers are implemented. In this work, RML models are directly parameterized by the image pixels, and are not composed of CLEAN-like components, nor convolved with a restoring beam (which can artificially degrade the resolution of the image).

Though CLEAN and RML are distinct image synthesis methods, both have been used successfully and complimentarily for ALMA protoplanetary disk observations \citep[e.g.][]{Carcamo_2018, Perez_2019, Yamaguchi_2020, Yamaguchi_2021, Casassus_2021, Zawadzki_2023}. In particular, \citet{Carcamo_2018, Carcamo_2019} introduced the \texttt{GPUVMEM} package which enabled GPU-accelerated RML imaging with a focus on ALMA protoplanetary disk imaging, establishing the use of efficient RML frameworks in the disk community. Utilizing multiple imaging methods is particularly worthwhile for testing the robustness of small or otherwise tenuous features within a disk, such as emission features likely caused by disk instabilities or embedded planets. Detecting these features in multiple independently synthesized images provides an additional layer of confidence that these features are real, as these features can be subtle and difficult to distinguish from spurious features arising from the imaging procedure rather than the data itself \citep[as done with the Event Horizon Telescope observations of M87, e.g.][]{Chael_2018, EHTCollab_2019}. For example, the imperfect subtraction of the instrument PSF during the CLEANing process could cause faint speckles to appear in the image, causing misinterpretations in a study attempting to directly image planets. Suppose a NKF appears in the source emission regardless of the imaging method used. In that case, the detected features are more likely real than an artifact stemming from a specific imaging choice. If detected features do not appear in independently synthesized images, then extra care should be taken to validate the feature before drawing conclusions about the presence of protoplanets or other disk processes. In order to fully characterize the physical and dynamical structures of the disk at high confidence, a comprehensive approach to imaging is essential. 

Artifacts may be introduced either through imperfect calibration or imaging procedures, the former of which would affect both the CLEAN and RML images. \citet{Loomis_exoALMA} describe steps taken to validate the exoALMA calibration process, and here we focus on improving confidence in the imaging results. We present the results of RML imaging applied to seven exoALMA sources which displayed prominent NKFs in their $^{12}$CO J=3-2 CLEAN images. The source selection was motivated by a preliminary visual search for kink-like substructures caused by planets, resulting in an initial sample of five disks: AA~Tau, J1615, J1842, LkCa~15, and SY~Cha. These are the same five sources for which \citet{Pinte_exoALMA} find evidence of embedded protoplanets and place constraints on the planet masses. To explore the effects of RML imaging on a wider variety of substructures, we add HD~135344B and J1604 to our sample, which both display large scale non-Keplerian arcs \citep[see Figure 1 in][]{Pinte_exoALMA}. For each of these seven sources we present full $^{12}$CO J=3-2, $^{13}$CO J=3-2, and CS J=7-6 image cubes, as well as comparisons to the CLEAN images for select channels of interest.

\section{Source Selection and Data Processing} \label{sec:data}

The exoALMA large program obtained data of 15 large and bright protoplanetary disks at moderate angular (maximum baselines up to 3697 m) and high spectral (up to 30.5 kHz, 26 m/s) resolution \citep{Teague_exoALMA, Loomis_exoALMA}. Here, we use CLEAN image cubes with 100 m/s channel spacings and $0\farcs15$ restoring beams as the fiducial image products for comparison with the RML image products; this is equivalent to the fiducial set of exoALMA images for $^{12}$CO J=3-2 and $^{13}$CO J=3-2, but with a finer channel spacing than the CS J=7-6 fiducial set of images (which has 200 m/s spacings). The full data calibration and CLEAN imaging procedures are detailed in \citet{Loomis_exoALMA}.

In this paper, we focus on select exoALMA targets which show clear NKFs in the $^{12}$CO J=3-2 emission imaged with \texttt{tclean}: protoplanetary disks around AA~Tau, HD~135344B, J1604, J1615, J1842, LkCa~15, and SY~Cha. For an in-depth analysis of the NKFs in the exoALMA sample, see \citet{Pinte_exoALMA}. To prepare the data for RML imaging, we first transformed each calibrated measurement set (i.e. the set of ($u$,$v$) points, complex visibilities, and visibility weights at each frequency) using the \texttt{CASA cvel2} task such that the data were in an LSRK reference frame. This must be done explicitly prior to exporting the data for RML imaging, as the reference frame correction is normally built into the \texttt{tclean} procedure. The data were regridded to a channel width of $0.1$ km/s, however, the regridding process is not necessary for RML imaging; we regridded the data to enable a direct comparison to the fiducial CLEAN images.

We use the RML imaging package \texttt{MPoL} for all imaging \citep{mpol_2023, Zawadzki_2023, mpol_2025}. We exported each of the regridded measurement sets (both the continuum-subtracted and non-continuum-subtracted $^{12}$CO J=3-2, $^{13}$CO J=3-2, and CS J=7-6 measurement sets for each disk) as arrays of complex visibilities for use with \texttt{MPoL}, and averaged the ungridded visibility data to grid cells in the visibility domain before imaging (equivalent to uniform weighting). The grid cells are specified in the image domain; in this study, all RML images are $1024 \times 1024$ pixels, with each pixel measuring $0\farcs0125$ across (half the size of the fiducial CLEAN pixels). This pixel scaling was chosen to ensure that the data with the highest angular resolutions (up to 0\farcs06) still have several pixels across a resolution element, while still capturing the full extent of the disk emission. We then defined a corresponding $1024 \times 1024$ Fourier grid and averaged the ungridded visibility data to the Fourier cells using a simple weighted average. We have verified that this gridding procedure does not introduce interpolation artifacts to the images at the SNR of the data.

\section{RML Techniques for Image Cubes} \label{sec:rml}

RML imaging aims to find the set of visibilities which maximizes the likelihood function
\begin{equation} \label{eqn:basic_likelihood}
    p\left(\mathbf{D} \mid \mathbf{I}\right),
\end{equation}
where $\mathbf{D}$ is a set of visibility data and $\mathbf{I}$ is a model image. In practice, we compute the negative log likelihood for computational efficiency. Assuming that the noise is normally distributed with standard deviation $\sigma$ and is uncorrelated across baselines, the negative natural logarithm of the likelihood function is 
\begin{equation}
    -\ln p\left(\mathbf{D} \mid \mathbf{I} \right) = N_{\rm{D}} \ln(\sqrt{2\pi}\sigma)+\frac{1}{2} \sum_{i}^{N_{\rm{D}}} \bigg| \frac{\rm{D}_{i}-\rm{V}_{i}(\mathbf{I})}{\sigma_{i}}\bigg|^{2},
\end{equation}
where $N_{D}$ is the number of complex visibilities in the dataset, $\rm{D}_{i}$ is a measured complex visibility at a $(u_{i},v_{i})$ point, and $\rm{V}_{i}$ is the predicted value of the model visibilities for the same $(u_{i},v_{i})$ generated from the model image. The rightmost term is equivalent to 1/2 of $\chi^{2}$ statistic, and the negative log likelihood can be written as
\begin{equation}
    L_{\mathrm{nll}}(\mathbf{I}) = -\ln p\left(\mathbf{D} \mid \mathbf{I} \right) = \frac{1}{2} \chi^{2}(\mathbf{D} \mid \mathbf{I}).
\end{equation}

The $L_{\mathrm{nll}}(\mathbf{I})$ notation is adapted from the machine learning community, where it is common to identify a function that can be minimized in order to obtain optimal model parameter values, known as a loss function \citep{Bishop_2006}. Though minimizing $L_{\mathrm{nll}}(\mathbf{I})$ is equivalent to maximizing the likelihood function, the resulting image may not be optimal due to the incomplete ($u,v$) sampling of the data. In other words, the ill-posed and underdetermined nature of radio interferometric imaging means that the maximum likelihood image cannot be uniquely determined. An image product can be improved from this base maximum likelihood image by incorporating one or more regularizers into the loss function; we adopt the same basic loss function as in 
\citet[hereafter \citetalias{Zawadzki_2023}]{Zawadzki_2023}, using a combination of entropy, sparsity, and total squared variation \citep[TSV,][]{Kuramochi_2018} regularization.

\subsection{The Loss Function}

The maximum entropy loss is defined as
\begin{equation}
    L_{\rm{ent}}=\frac{1}{\zeta} \sum_{i} I_{i} \ln I_{i},
\end{equation}
where $I_i$ is the intensity of an image pixel. Here, $\zeta$ is a normalization factor which we set equal to the total flux of the CLEAN image. Maximum entropy regularization restricts pixel values to positive non-zero values and favors uniformity in the image, making it useful for identifying emission features \citep{Narayan_1986, Gull_1978, Hoegbom_1979}.

The sparsity loss is defined as
\begin{equation}
    L_{\rm{spa}}=\sum_{i}\left|I_{i}\right| 
\end{equation}
and is derived from the least absolute shrinkage and selection operator \citep[lasso, ][]{Tibshirani_1996}.
Sparsity regularization uses the $L_{1}$ norm to suppress the amplitudes of low-intensity pixels, promoting an image that is a sparse collection of non-zero valued pixels. This often has the effect of greatly reducing background noise in the image and improving image resolution \citep{Honma_2014}. 

Finally, the TSV loss is defined as
\begin{equation}
L_{\rm{TSV}}=\sum_{l, m}\left(I_{l+1, m}-I_{l, m}\right)^{2}+\left(I_{l, m+1}-I_{l, m}\right)^{2},
\end{equation}
where $l$ and $m$ are indices over pixels corresponding to right ascension and declination, respectively.
TSV regularization takes the contiguity of image pixels into account, favoring sharp changes in intensity when needed and relatively smooth areas elsewhere.

Each of the above regularizers has an adjustable coefficient $\lambda$ which determines the strength of that regularizer. Since each $\lambda$ affects the values of the model parameters ($\mathbf{I}$) without itself being a free parameter in the model, it is known as a hyperparameter. The total loss function is the sum of the likelihood function and all regularizing terms,
\begin{multline} \label{eqn:loss}
    L(\mathbf{I}) = L_{\mathrm{nll}}(\mathbf{I}) + \lambda_{\mathrm{ent}} L_{\mathrm{ent}}(\mathbf{I}) + \lambda_{\mathrm{spa}} L_{\mathrm{spa}}(\mathbf{I}) + \\ \lambda_{\mathrm{TSV}} L_{\mathrm{TSV}}(\mathbf{I}).
\end{multline}

\subsection{Cross-Validation for Multi-Channeled Data} \label{sec:cvmethods}

\begin{figure*}
\centering
\includegraphics[width=\linewidth]{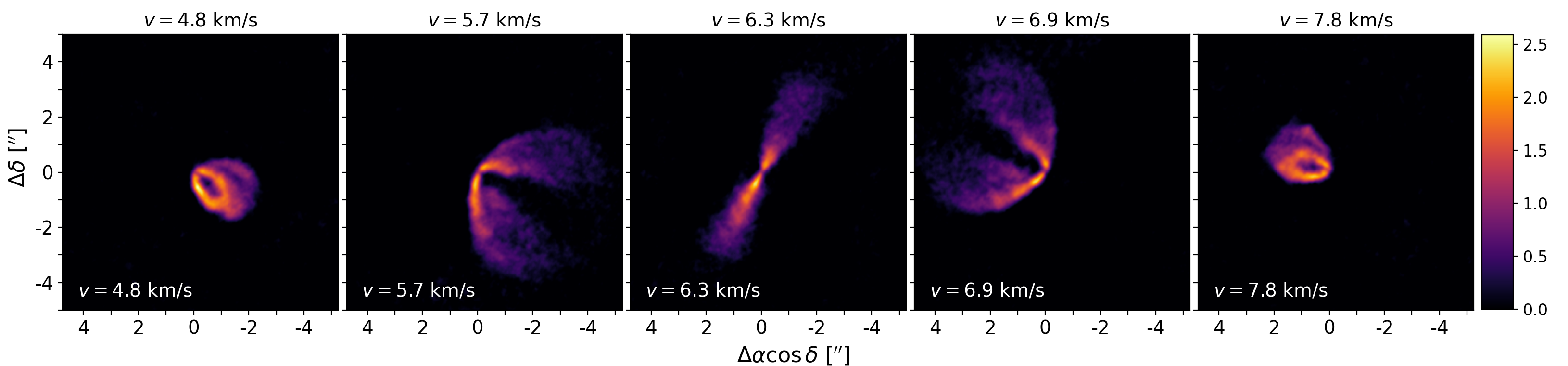}
\caption{The five channels of LkCa~15 selected for thorough CV testing (RML images pictured). Channels were selected arbitrarily across the range of velocity space where gas emission is prominent in all three molecular lines, testing channels that varied in the spatial extent and general morphology of the emission. We intentionally include a channel with a prominent NKF at $v=6.9$ km/s. The color bar has units of Jy/arcsec$^2$.}
\label{fig:cvchans}
\end{figure*}

Cross-validation methods are well-established in astronomy as a quantitative procedure for obtaining hyperparameter values \citep[e.g.][]{Akiyama_2017a, Akiyama_2017b, Yamaguchi_2020, Yamaguchi_2021, Yamaguchi_2024, Aizawa_2020, Zawadzki_2023}. Following the suggestions in \citetalias{Zawadzki_2023}, we use random cell cross-validation (CV) to identify the optimal hyperparameters ($\lambda_{\mathrm{ent}}$, $\lambda_{\mathrm{spa}}$, and $\lambda_{\mathrm{TSV}}$) for imaging. Random cell CV works by partitioning visibility grid cells into $K$ pairs of training and testing sets randomly, drawing cells without replacement with the minor exception that grid cells with the highest 1\% of gridded weight values are included in each subset in order to avoid numerical instabilities that delay or prevent convergence. After defining the loss function, an RML model is generated using the training data. This model is then compared to the withheld testing data to quantify the predictive power of the RML model and obtain a CV score,
\begin{equation}
    \rm{CV} = \sum^{K}_{k} \chi^{2}(\mathbf{D}_{k} \mid \mathbf{I}_{\rm train}),
\end{equation}
where $\mathbf{D}_{k}$ is the test dataset, and $\mathbf{I}_{\rm train}$ is the model image that minimizes the loss function for the training data $\mathbf{D} - \mathbf{D}_{k}$. We use $K=10$, which has been shown to effectively balance bias and variance in the parameter error estimates \citep{Breiman_1992, Kohavi_1995, Molinaro_2005}.

Previous work on RML imaging for protoplanetary disk observations focused primarily on continuum images \cite[e.g][]{Carcamo_2018, Casassus_2019b, Perez_2019, Perez_2020, Yamaguchi_2020, Yamaguchi_2021, Yamaguchi_2024, Zawadzki_2023}, however, extending the application of RML methods to velocity-resolved gas observations presents a new set of challenges. While some studies have used RML methods for imaging the gas emission in protoplanetary disks \citep[e.g.][]{Carcamo_2018}, only a few present velocity-resolved RML channel maps \citep{Casassus_2021, Casassus_2022}. As a result, there is not yet a firm consensus on best practices for applying RML techniques to multi-channeled data like the observations presented here. In particular, it is unclear whether CV must be performed on a channel-by-channel basis, or whether CV results from a representative channel can be applied to the rest of (or at least a subsection of) the full data cube. It is also unclear whether optimal hyperparameter values are comparable for different observations or data products of the same source (e.g. $^{12}$CO J=3-2 vs. $^{13}$CO J=3-2, or continuum-subtracted vs. non-continuum-subtracted data). The CV process is by far the most computationally expensive component of RML imaging, so eliminating unnecessary hyperparameter tuning results in significant time and resource savings when working with a large sample like the exoALMA disks.

\begin{table}[]
\centering
  \caption{Top: The tested range of regularizer hyperparameter values ($\lambda_{\mathrm{ent}}$, $\lambda_{\mathrm{spa}}$, and $\lambda_{\mathrm{TSV}}$). Botton: Optimal entropy, TSV, and sparsity $\lambda$ values resulting from random cell CV on various LkCa~15 data. The full CV process was conducted on 1) $^{12}$CO J=3-2, $^{13}$CO J=3-2, and CS J=7-6 data, 2) both the continuum-subtracted and non-continuum-subtracted observations, and 3) five individual channels (100 m/s channel width) across the data cubes displaying different emission morphologies, including the channel with the apparent NKF at 6.9 km/s. The optimal $\lambda$ values resulting from CV show little variation across channels or lines, suggesting that CV on a single representative channel is sufficient for setting hyperparameter values for an entire data cube or multiple similar observations of the same source, greatly reducing the computational burden of RML imaging across large datasets.}
  \label{tab:cv_results}
\begin{tabular}{lcllll}
\hline
\multicolumn{6}{c}{tested hyperparameter values} \\ \hline
\multicolumn{1}{l}{$\lambda_{\rm{ent}}$} & \multicolumn{5}{l}{{[}0, 8e-8, 2e-7, 8e-7, 2e-6, 8e-6, 2e-5{]}} \\
\multicolumn{1}{l}{$\lambda_{\rm{TSV}}$} & \multicolumn{5}{l}{{[}5e-5, 1e-4, 5e-4, 1e-3, 5e-3, 1e-2{]}} \\
\multicolumn{1}{l}{$\lambda_{\rm{spa}}$} & \multicolumn{5}{l}{{[}5e-6, 1e-5, 5e-5{]}} \\ \hline
line & contsub? & channel & $\lambda_{\rm{ent}}$ & $\lambda_{\rm{TSV}}$ & $\lambda_{\rm{spa}}$ \\ \hline
$^{12}$CO J=3-2& Yes & 6.9 km/s & 0   & 5e-4 & 1e-5 \\
$^{12}$CO J=3-2& Yes & 6.3 km/s & 0   & 5e-4 & 1e-5 \\
$^{12}$CO J=3-2& Yes & 4.8 km/s & 0   & 1e-4 & 1e-5 \\
$^{12}$CO J=3-2& No  & 6.9 km/s & 0   & 5e-4 & 1e-5 \\
$^{12}$CO J=3-2& No  & 6.3 km/s & 0   & 5e-4 & 1e-5 \\
$^{12}$CO J=3-2& No  & 4.8 km/s & 8e-8& 1e-4 & 1e-5 \\
$^{13}$CO J=3-2& Yes & 6.9 km/s & 0   & 5e-4 & 1e-5 \\
$^{13}$CO J=3-2& Yes & 6.3 km/s & 0   & 5e-4 & 1e-5 \\
$^{13}$CO J=3-2& Yes & 4.8 km/s & 0   & 5e-4 & 1e-5 \\
CS        J=7-6& Yes & 6.9 km/s & 0   & 5e-4 & 1e-5 \\
CS        J=7-6& Yes & 6.3 km/s & 0   & 5e-4 & 1e-5 \\
CS        J=7-6& Yes & 4.8 km/s & 0   & 1e-3 & 1e-5 
\end{tabular}
\end{table}

To test this we performed CV on the LkCa~15 data, evaluating a grid of entropy, TSV, and sparsity hyperparameter values spanning several orders of magnitude. We included the $^{12}$CO J=3-2, $^{13}$CO J=3-2, and CS J=7-6 observations and tested both the continuum-subtracted and non-continuum-subtracted observations across multiple different channels sampling different parts of the velocity space. Our findings are summarized in Table \ref{tab:cv_results}, showing results from a subset of our 30 CV tests. In total, we tested 5 non-adjacent channels for each line (both continuum-subtracted and non-continuum-subtracted). These channels are presented in Figure \ref{fig:cvchans}, and were selected to probe different spatial extents and general morphologies of the emission.

The optimal hyperparameter values tended to remain nearly constant throughout a given image cube, with only slight variations from channel to channel or no variation at all. We verified that these minor variations did not significantly change the emission morphologies in the resulting images. Optimal hyperparameters resulting from the CV procedures also remained largely consistent between the continuum-subtracted and non-continuum-subtracted data. Lastly, hyperparameter values remained similar across the $^{12}$CO J=3-2, $^{13}$CO J=3-2, and CS J=7-6 data, despite often having substantial differences in emission morphology and intensity.

These results suggest that hyperparameter values are more dependent on the observational parameters (e.g. interferometer baselines and on-source integration time, corresponding to the resolution and sensitivity of the data) rather than the intensities or specific morphological features across different molecular lines. Thus, it is possible to exploit similarities between data to significantly reduce the computational burden of CV for RML hyperparameter tuning. A typical CV procedure for one combination of hyperparameters can take $30$ minutes on a NVIDIA P100 GPU; with a comprehensive grid of hyperparameter values, the process requires tens of hours (1-2 days) on a single GPU. If the CV-tuned hyperparameters showed substantial variation within and across the cubes, obtaining hyperparameter values with CV would quickly become impractical for large datasets like the exoALMA data.

Given these results, we followed the RML workflow laid out in \citetalias{Zawadzki_2023}, except adapted for working with multi-channeled data:
\begin{enumerate}
    \item For each source, we selected a single channel to use for CV by examining the fiducial continuum-subtracted $^{12}$CO J=3-2 data CLEAN images. For sources with a spatially small NKF localized to a few channels, we used one of the channels where the NKF is visible in the fiducial CLEAN images. For sources with large scale NKFs spanning many channels, we selected a representative channel where the NKFs are clearly seen in the fiducial CLEAN images. Following the extensive CV testing summarized in Table \ref{tab:cv_results}, we do not expect the choice of channel at this step to have a significant impact on the final RML images.
    \item We performed 10-fold CV on the continuum-subtracted $^{12}$CO J=3-2 data at this channel for the range of hyperparameter values ($\lambda_{\mathrm{ent}}$, $\lambda_{\mathrm{TSV}}$, and $\lambda_{\mathrm{spa}}$) presented in Table \ref{tab:cv_results}.
    \item We applied the hyperparameter values that minimized the CV score at this channel to our loss function (Equation \ref{eqn:loss}). We used this loss function for all channels in the cube.
    \item With the loss function for each source now specified with the optimized hyperparameters, we synthesized full RML image cubes for the continuum-subtracted and non-continuum-subtracted $^{12}$CO J=3-2, $^{13}$CO J=3-2, and CS J=7-6 data.
\end{enumerate}

\begin{table}[]
\centering
  \caption{Hyperparameter values resulting from 10-fold CV on the continuum-subtracted $^{12}$CO J=3-2 data for each source. These values were obtained from running CV on a single channel of interest (100 m/s channel width), then applied to the rest of the cube.} \label{tab:cv_summary}
\hspace*{-1.5cm} 
\begin{tabular}{lcccc}
Source & Channel (km/s) & $\lambda_{\rm{ent}}$ & $\lambda_{\rm{TSV}}$ & $\lambda_{\rm{spa}}$ \\ \hline
AA~Tau      & $v=7.9$  & 0     & 5e-5  & 1e-5  \\
HD~135344B  & $v=6.6$  & 0     & 5e-5  & 5e-5  \\
J1604       & $v=4.5$  & 0     & 1e-4  & 5e-5  \\
J1615       & $v=3.7$  & 8e-6  & 5e-4  & 5e-5  \\
J1842       & $v=5.1$  & 8e-6  & 5e-5  & 5e-5  \\
LkCa~15     & $v=6.9$  & 0     & 5e-4  & 1e-5  \\
SY~Cha      & $v=3.6$  & 0     & 5e-4  & 1e-5 
\end{tabular}
\end{table}

The CV results are summarized in Table~\ref{tab:cv_summary}. Five of the seven sources yielded optimal $\lambda_{\rm{ent}}$ values of zero, effectively ``turning off'' entropy regularization and indicating that TSV and sparsity regularization alone yielded the model with the best predictive power between the training and testing data during CV. 
Other recent studies have also obtained high quality protoplanetary disk images using only a combination of TSV and sparsity regularization \citep[e.g.][]{Yamaguchi_2024}. The two remaining disks, J1615 and J1842, favor a modest degree of entropy regularization and each have an optimal $\lambda_{\rm{ent}}$ of $8\times10^{-6}$. For all seven disks, the optimal $\lambda_{\rm{TSV}}$ values range from $5\times10^{-5}$ to $5\times10^{-4}$, and the optimal $\lambda_{\rm{spa}}$ values range from $1\times10^{-5}$ to $5\times10^{-5}$.

We note that it is also possible to continue tuning hyperparameters by hand after the CV process. However, deviating from the set of hyperparameters which minimize the CV score introduces human bias, as hand tuning is simply adjusting the hyperparameters manually to obtain a more desirable image appearance (evaluated by eye rather than a quantitative metric like a CV score; see the discussion on hyperparameter tuning in \citetalias{Zawadzki_2023}). To treat the exoALMA sample more uniformly, we do not take this optional tuning step and use strictly the hyperparameter values which minimized the CV score for each source.

We also find that the optimal hyperparameter values remain similar across different exoALMA sources, however, there are small differences between most sources (Table \ref{tab:cv_summary}). We do not automatically apply any results from one source to another, conducting CV independently for each disk. As optimal hyperparameter values appear to be driven primarily by the ($u,v$) coverage and sensitivity of the observations, CV results from one target can be used as a starting point for the rest of the sample in large, relatively uniform programs like exoALMA. Given that only two disks (LkCa~15 and SY~Cha) had identical optimal hyperparameters, however, we recommend performing CV at least once per source. 

This apparent dependence on ($u,v$) coverage and sensitivity also means that the simple CV procedure used here may not be as effective for observations with less uniform ($u,v$) coverage or lower signal-to-noise ratios. For applications to such data, it may be necessary to repeat the CV process multiple times with different visibility partitioning to further validate and avoid artificially skewing the hyperparameter values based on the random partitioning into training and testing sets. In this case, it would also be useful to test the performance of CV on simulated data with comparably sparse or non-uniform ($u,v$) coverage to better characterize the behavior of CV in this low sensitivity regime.


\section{Results} \label{sec:results}

\begin{figure}
\centering
\includegraphics[width=\linewidth]{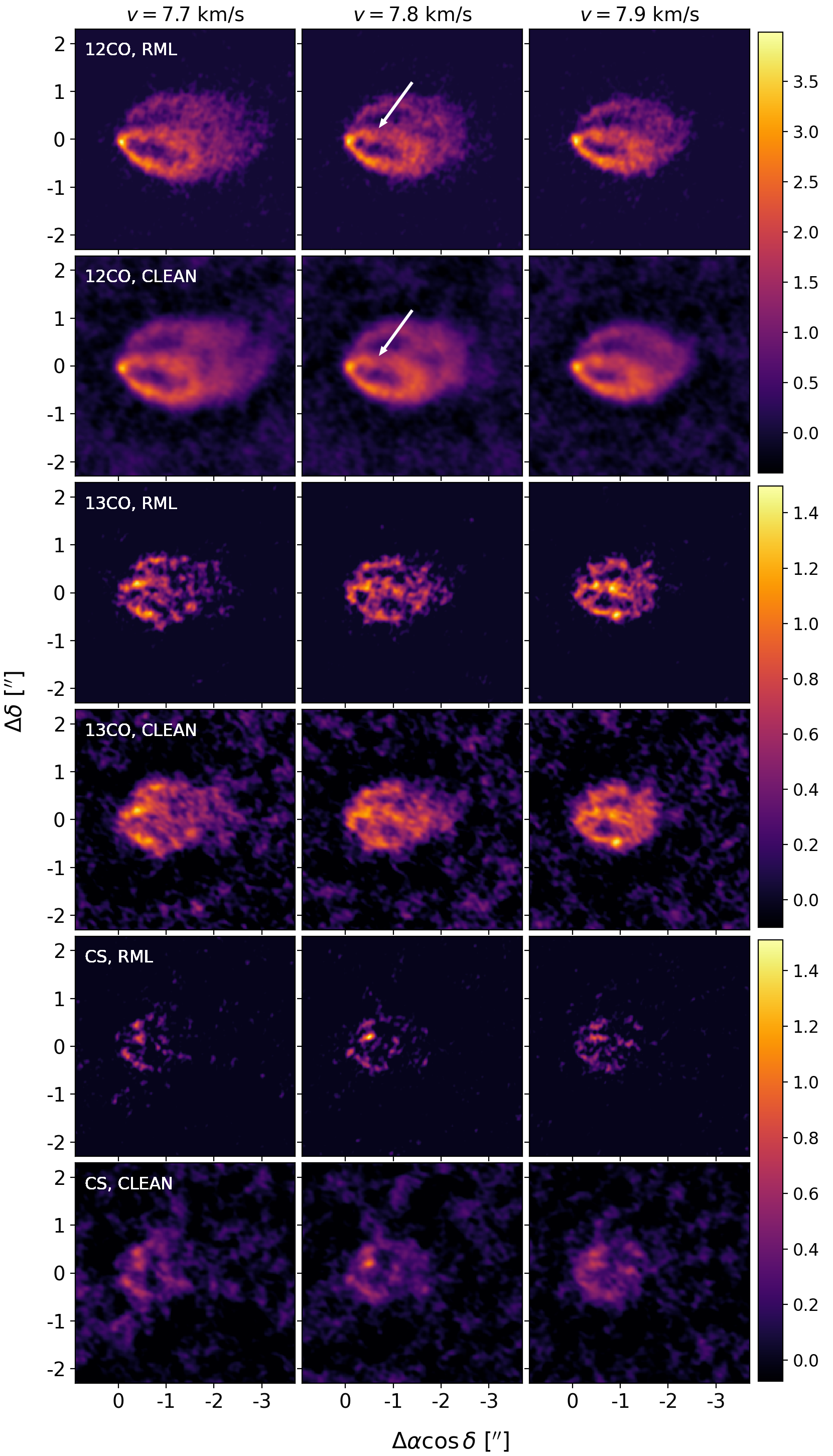}
\caption{RML and CLEAN comparison for continuum-subtracted observations of AA~Tau in $^{12}$CO J=3-2, $^{13}$CO J=3-2, and CS J=7-6. Shown are three adjacent channels, centered at $v=7.8$ km/s where the $^{12}$CO J=3-2 non-Keplerian feature appears most prominently. All 6 panels for each molecular line (the three RML panels and three CLEAN panels) are plotted on the same color scale, and the color bars for each molecule are in units of Jy/arcsec$^{2}$. The white arrows indicate the NKF which can be seen in the $^{12}$CO J=3-2 emission; while the feature is present in all three channels shown, we only place the arrows in one RML image and one CLEAN image to highlight the location of the feature.}
\label{fig:aatau}
\end{figure}

\begin{figure}
\centering
\includegraphics[width=\linewidth]{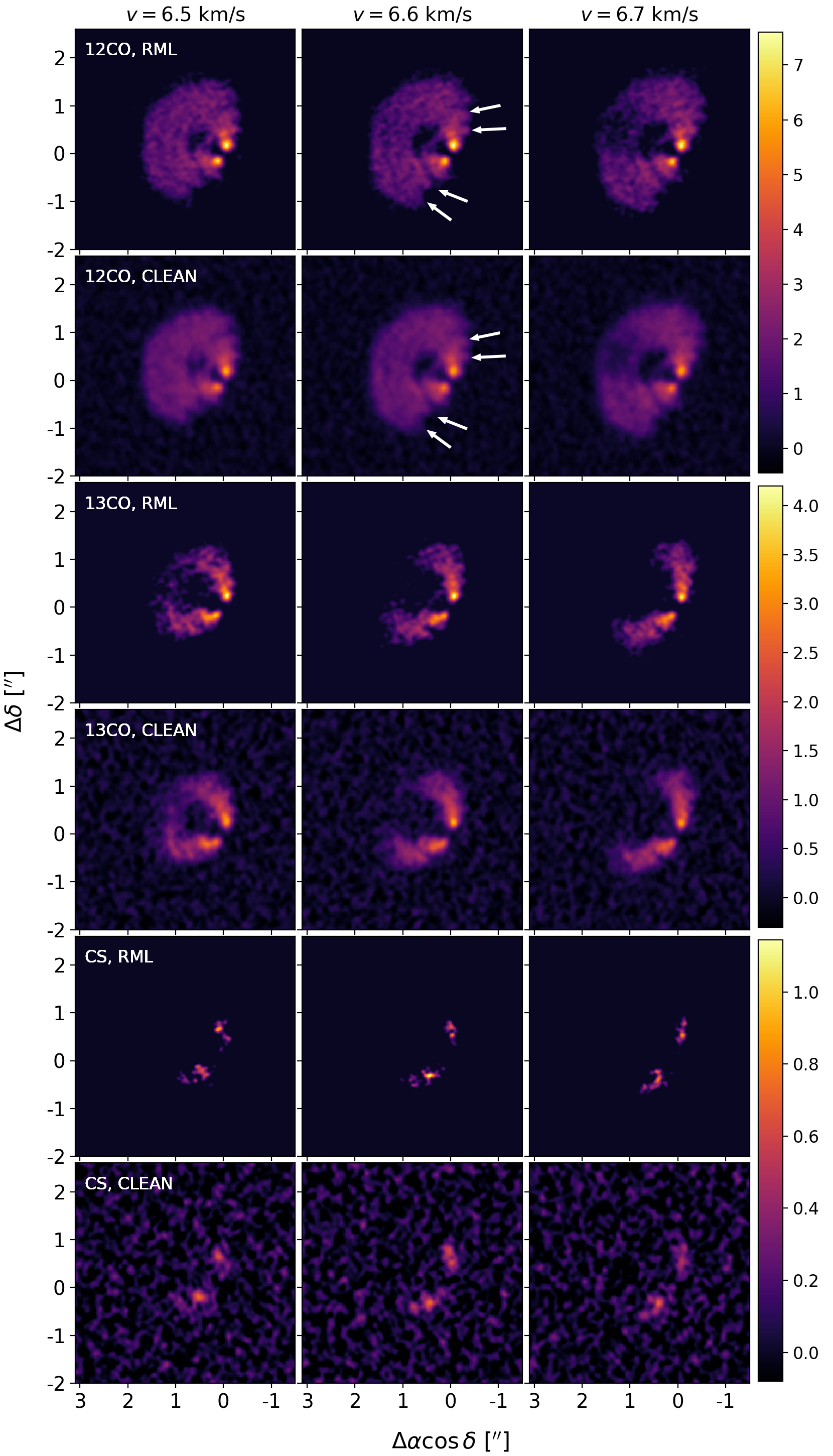}
\caption{Same as Figure \ref{fig:aatau}, but for HD~135344B. As HD~135344B displays large NKFs that persist throughout the cube, we show three adjacent and representative channels centered at $v=6.6$ km/s. }
\label{fig:hd135344b}
\end{figure}

\begin{figure}
\centering
\includegraphics[width=\linewidth]{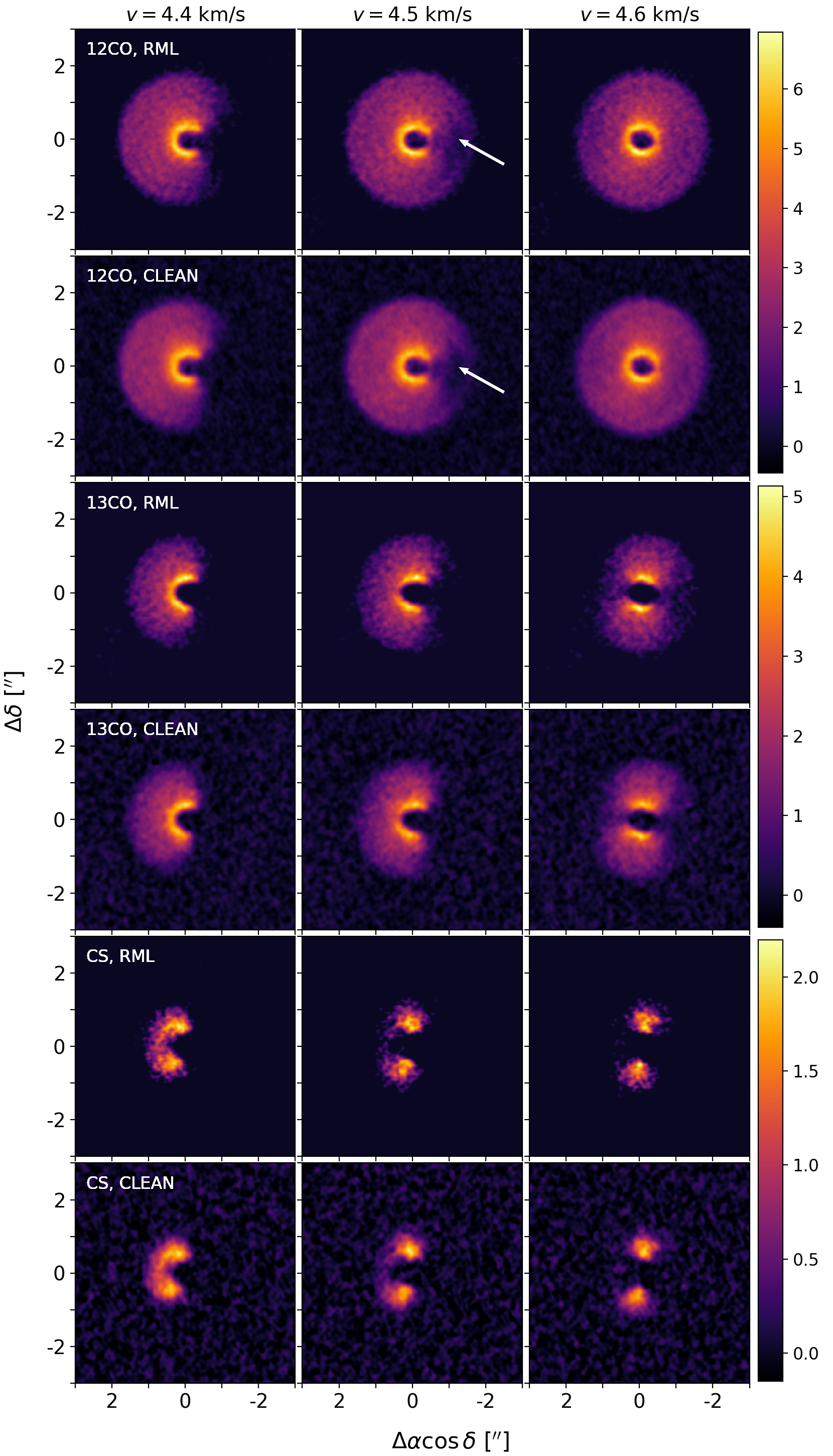}
\caption{Same as Figure \ref{fig:aatau}, but for J1604. As J1604 displays large NKFs that persist throughout the cube, we show three adjacent and representative channels centered at $v=4.5$ km/s.}
\label{fig:j1604}
\end{figure}

\begin{figure}
\centering
\includegraphics[width=\linewidth]{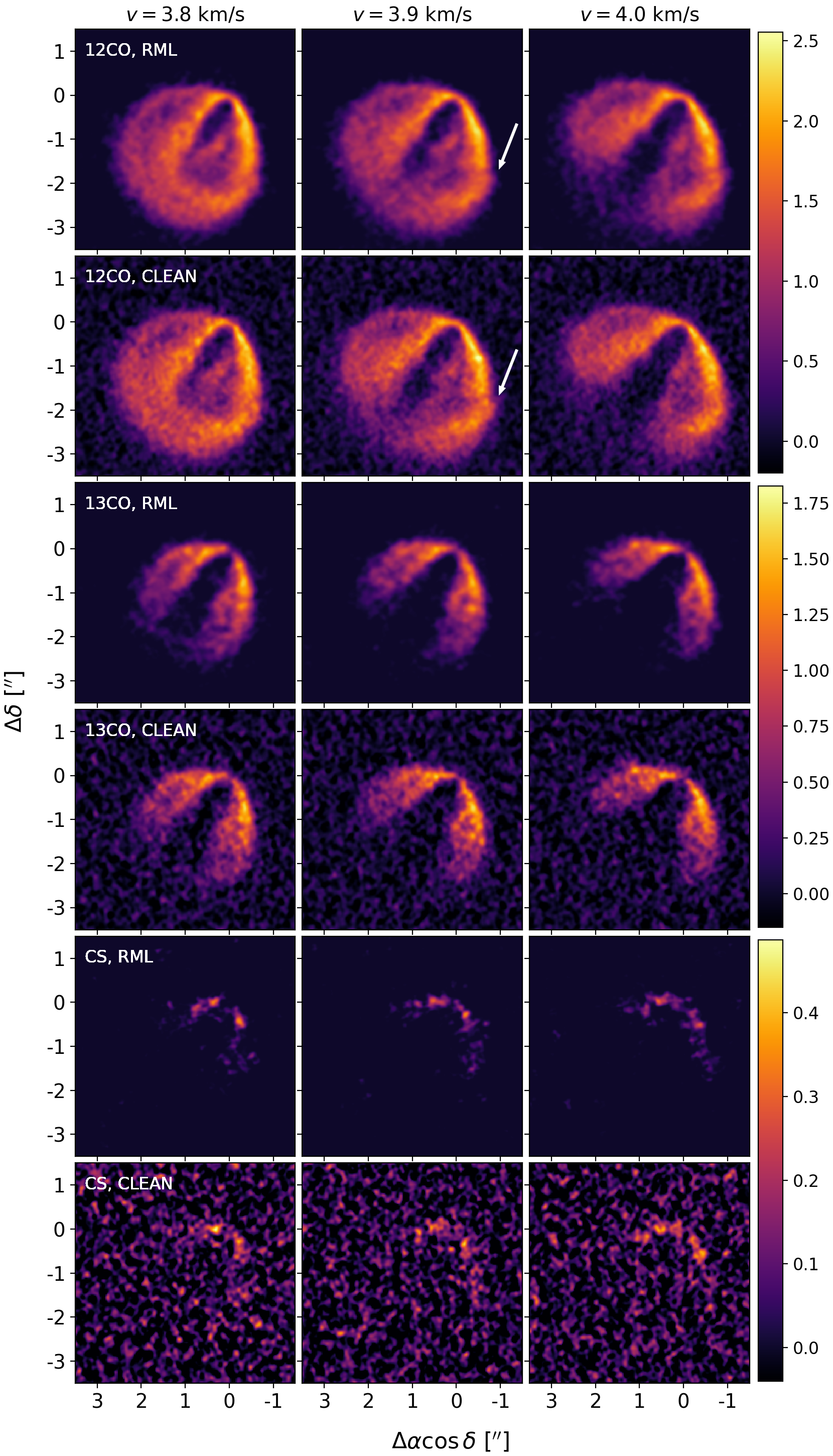}
\caption{Same as Figure \ref{fig:aatau}, but for J1615. Shown are three adjacent channels, centered at $v=3.9$ km/s where the $^{12}$CO J=3-2 non-Keplerian feature appears most prominently.}
\label{fig:j1615}
\end{figure}

\begin{figure}
\centering
\includegraphics[width=\linewidth]{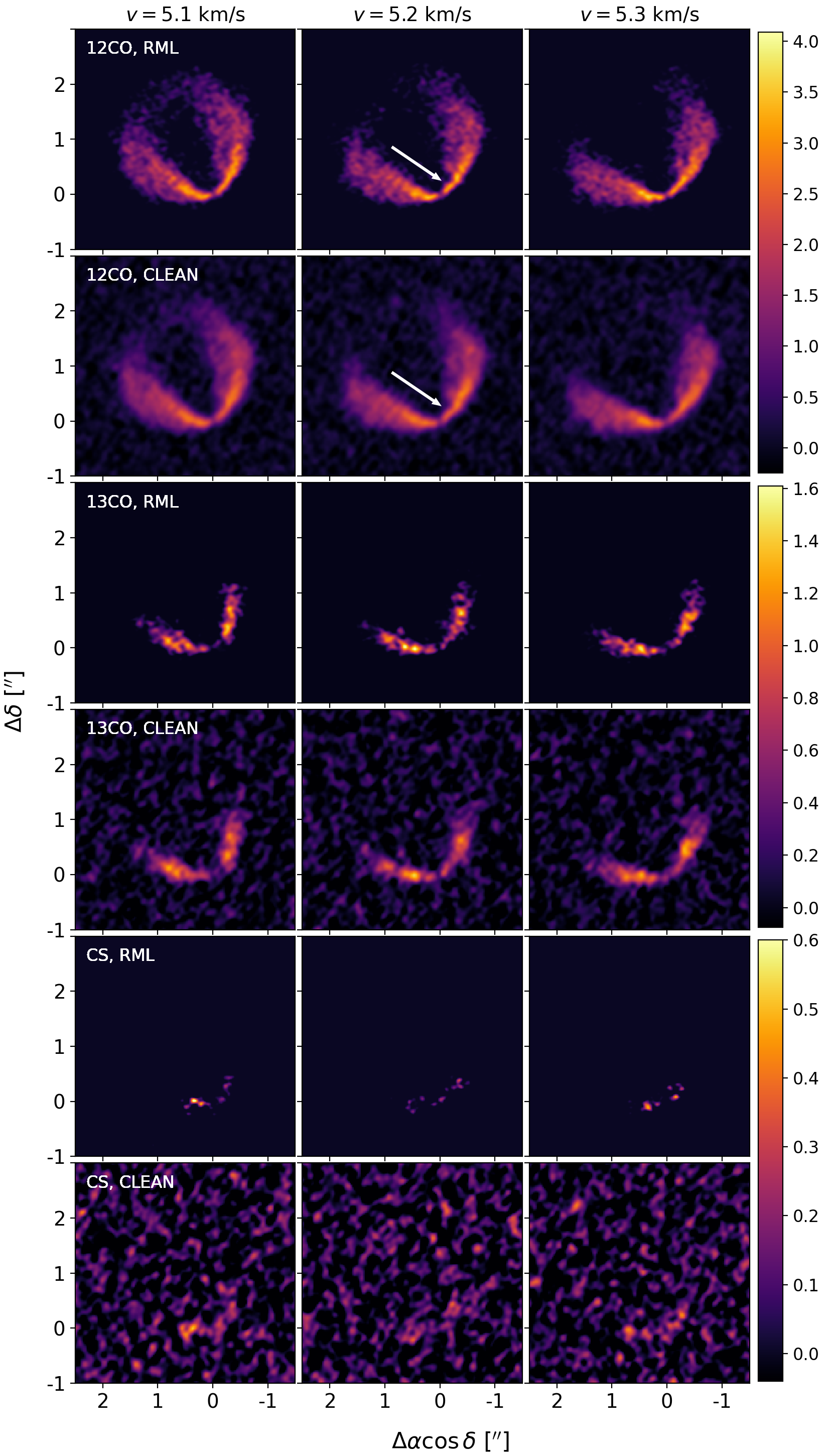}
\caption{Same as Figure \ref{fig:aatau}, but for J1842. Shown are three adjacent channels, centered at $v=5.2$ km/s where the $^{12}$CO J=3-2 non-Keplerian feature appears most prominently.}
\label{fig:j1842}
\end{figure}

\begin{figure}
\centering
\includegraphics[width=\linewidth]{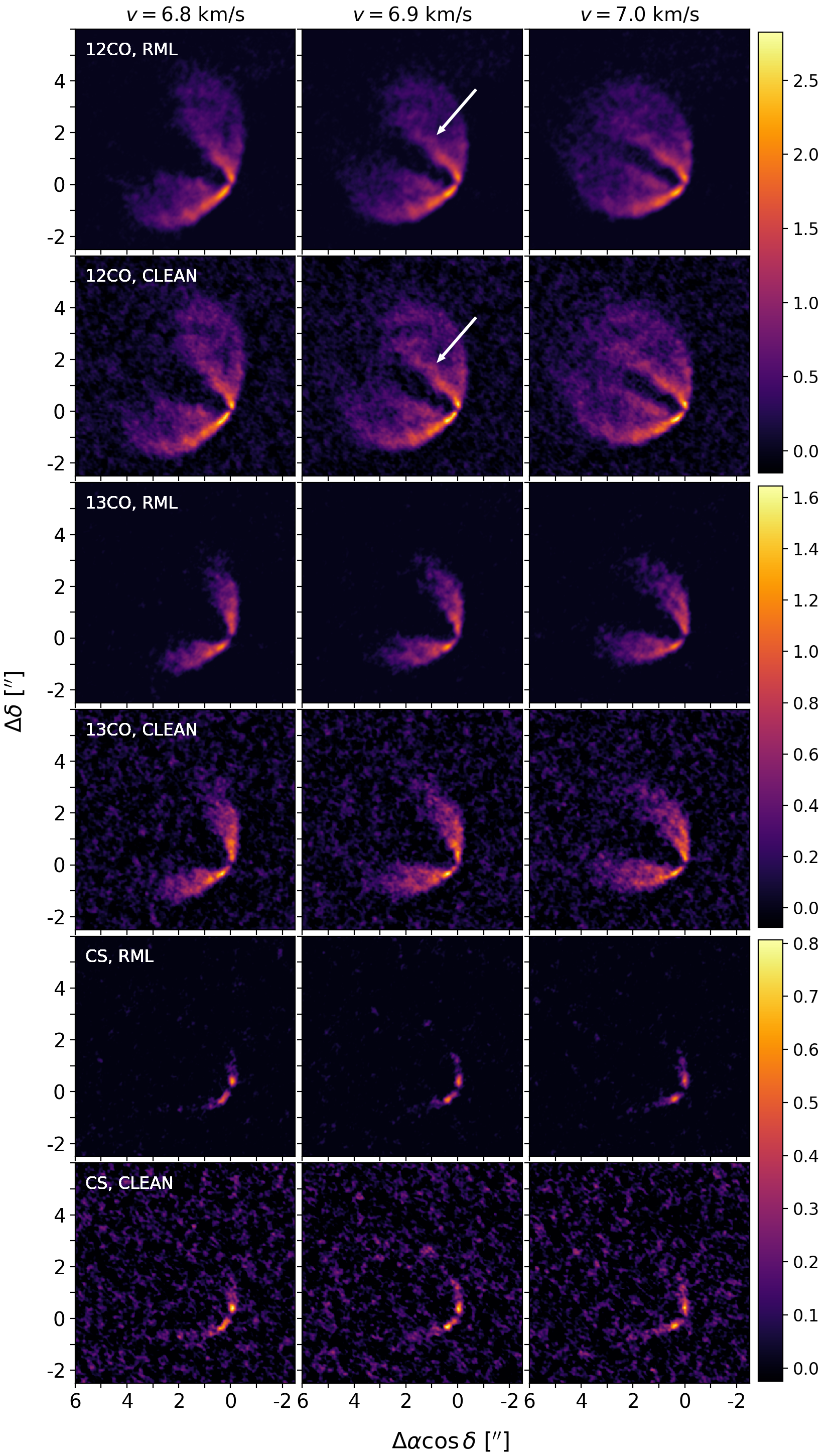}
\caption{Same as Figure \ref{fig:aatau}, but for LkCa~15. Shown are three adjacent channels, centered at $v=6.9$ km/s where the $^{12}$CO J=3-2 non-Keplerian feature appears most prominently.}
\label{fig:lkca15}
\end{figure}

\begin{figure}
\centering
\includegraphics[width=\linewidth]{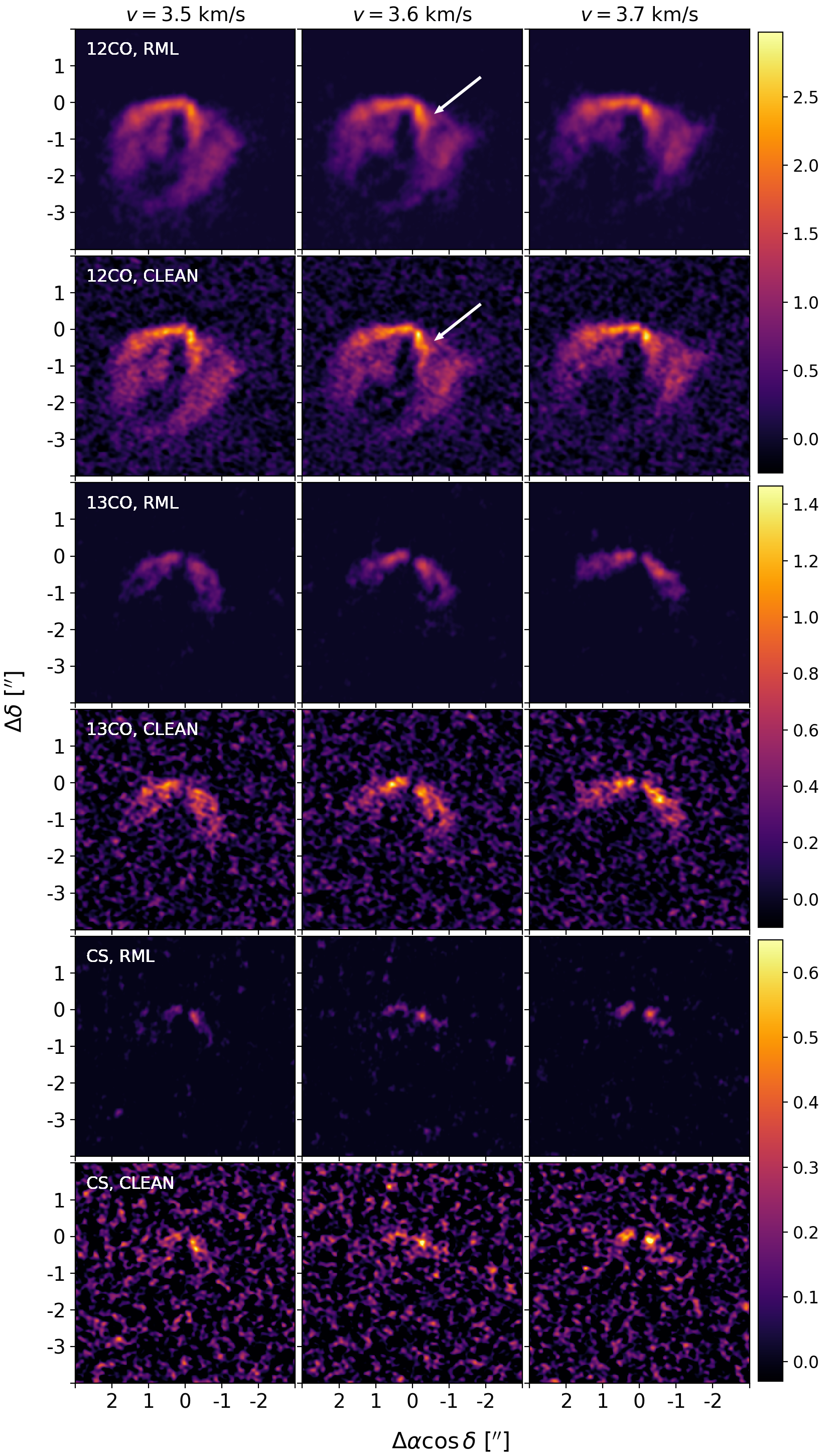}
\caption{Same as Figure \ref{fig:aatau}, but for SY~Cha. Shown are three adjacent channels, centered at $v=3.6$ km/s where the $^{12}$CO J=3-2 non-Keplerian feature appears most prominently.}
\label{fig:sycha}
\end{figure}

\begin{figure}
\centering
\includegraphics[width=0.867\linewidth]{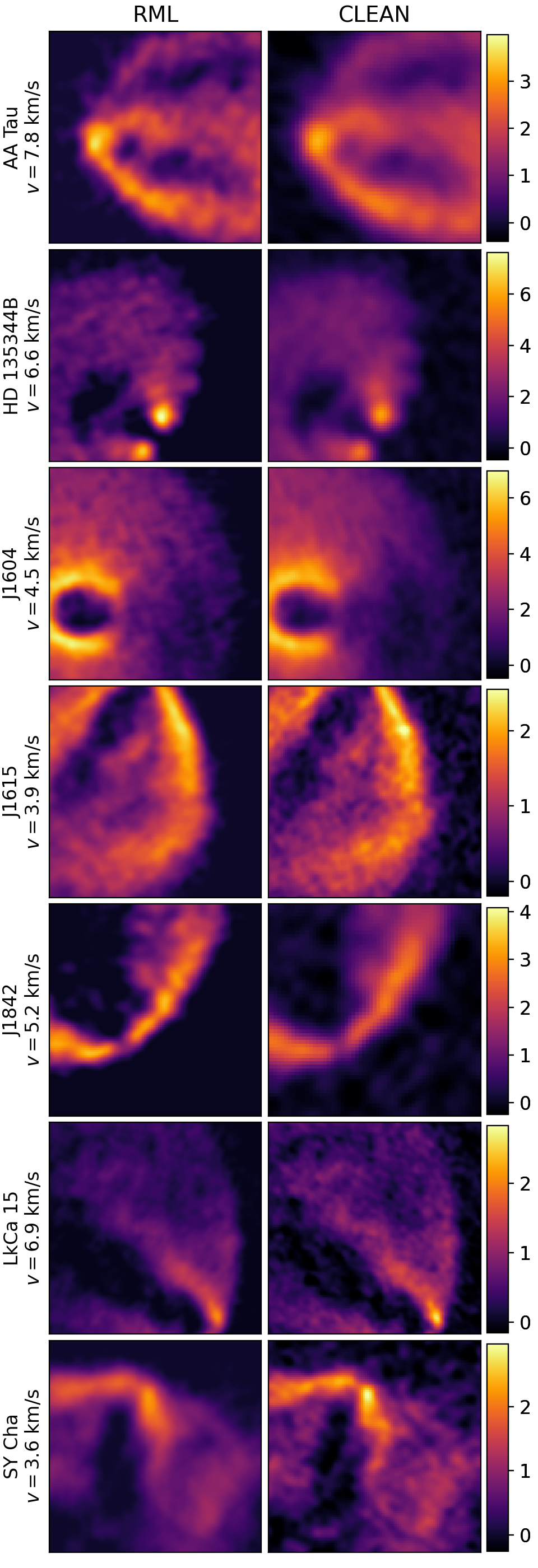}
\caption{A zoomed-in view of the $^{12}$CO J=3-2 NKFs present in the RML (left) and CLEAN (right) images for each source. The color bars are in units of Jy/arcsec$^{2}$.}
\label{fig:zoomkink}
\end{figure}

Following the recommendations from \citetalias{Zawadzki_2023} and incorporating the new findings for spectral line data described in Section \ref{sec:rml}, we generated continuum-subtracted and non-continuum-subtracted RML image cubes of the $^{12}$CO J=3-2, $^{13}$CO J=3-2, and CS J=7-6 emission with 100 m/s channel spacing for 7 exoALMA targets. Each source is summarized with a figure comparing the RML and CLEAN images (Figures \ref{fig:aatau}-\ref{fig:sycha}). Each figure displays 3 channels, centered on a channel where the NKF(s) appear prominently in the $^{12}$CO J=3-2 data \citep[see][]{Pinte_exoALMA}. Figure \ref{fig:zoomkink} displays the most prominent NKFs in the continuum-subtracted $^{12}$CO J=3-2 emission for each source, showing the difference between the RML and CLEAN images with a zoomed-in look at the NKFs. The RML images presented in these figures were made with the optimal hyperparameter values presented in Table \ref{tab:cv_summary} following the procedure described in Section \ref{sec:cvmethods}.

\subsection{AA~Tau}
The CLEAN images of AA~Tau show a NKF in the $^{12}$CO J=3-2 emission centered at roughly $v=7.8$ km/s and visible in multiple adjacent channels. Figure \ref{fig:aatau} shows a comparison of the RML and CLEAN images of AA~Tau for the continuum-subtracted $^{12}$CO J=3-2, $^{13}$CO J=3-2, and CS J=7-6 data. All of the RML images in Figure \ref{fig:aatau} were made with $\lambda_{\rm{ent}} = 0$, $\lambda_{\rm{TSV}} = 5\times 10^{-5}$, and $\lambda_{\rm{spa}} = 1\times 10^{-5}$.

The RML images of AA~Tau appear sharper than the corresponding CLEAN images. All the features appearing in the CLEAN images are recovered in the corresponding RML images, except the RML images have more uniform background emission due to the sparsity regularization. This can also be thought of as background noise suppression through regularization. The NKF is visible in both the RML and CLEAN $^{12}$CO J=3-2 images, and is consistent with a 2 $\rm{M}_{\rm{Jup}}$ planet embedded in the disk \citep{Pinte_exoALMA}. This feature is not as apparent in the $^{13}$CO J=3-2 or CS J=7-6 images, but there are bright clumps present in these channels at the rough location of the NKF.

\subsection{HD~135344B}
The CLEAN images of HD~135344B show NKFs consistently throughout the $^{12}$CO J=3-2 emission. Figure \ref{fig:hd135344b} shows a comparison of the RML and CLEAN images of HD~135344B for the continuum-subtracted $^{12}$CO J=3-2, $^{13}$CO J=3-2, and CS J=7-6 data. Though the NKFs in this disk are not localized to only a handful of channels like planet-driven NKFs, we select three adjacent channels centered at $v=6.6$ km/s for illustrative purposes. All of the RML images in Figure \ref{fig:hd135344b} were made with $\lambda_{\rm{ent}} = 0$, $\lambda_{\rm{TSV}} = 5\times 10^{-5}$, and $\lambda_{\rm{spa}} = 5\times 10^{-5}$.

The RML images of HD~135344B appear sharper than the corresponding CLEAN images, and have more uniform background emission due to the sparsity regularization. The RML images consistently reproduce the large scale NKFs seen in the CLEAN images.

\subsection{J1604}
The CLEAN images of J1604 show NKFs consistently throughout the $^{12}$CO J=3-2 emission. Figure \ref{fig:j1604} shows a comparison of the RML and CLEAN images of J1604 for the continuum-subtracted $^{12}$CO J=3-2, $^{13}$CO J=3-2, and CS J=7-6 data. Though the NKFs in this disk are not localized to only a handful of channels like planet-driven NKFs, we select three adjacent channels centered at $v=4.5$ km/s for illustrative purposes. All of the RML images in Figure \ref{fig:j1604} were made with $\lambda_{\rm{ent}} = 0$, $\lambda_{\rm{TSV}} = 1\times 10^{-4}$, and $\lambda_{\rm{spa}} = 5\times 10^{-5}$.

Here, the RML images look extremely similar to the fiducial CLEAN images, with only a small degree of added sharpness which appears more obvious in the fainter $^{13}$CO J=3-2 and CS J=7-6 emission. Features like the smoothness of the emission, edge sharpness, and presence of NKFs appear to be fully reproduced in the RML images. Background noise is suppressed in the RML images due to sparsity regularization.

\subsection{J1615}
The CLEAN images of J1615 show a NKF in the $^{12}$CO J=3-2 emission centered at roughly $v=3.9$ km/s. Figure \ref{fig:j1615} shows a comparison of the RML and CLEAN images of J1615 for the continuum-subtracted $^{12}$CO J=3-2, $^{13}$CO J=3-2, and CS J=7-6 data. All of the RML images in Figure \ref{fig:j1615} were made with $\lambda_{\rm{ent}} = 8\times 10^{-6}$, $\lambda_{\rm{TSV}} = 5\times 10^{-4}$, and $\lambda_{\rm{spa}} = 5\times 10^{-5}$.

While the RML images appear slightly more blurred than the CLEAN images for this target, the NKF remains visible in multiple channels. This feature is consistent with a 2 $ M_{\rm{Jup}}$ planet embedded in the disk \citep{Pinte_exoALMA}. Incorporating the additional step of hand-tuning hyperparameter values after CV may result in a sharper RML image. The largest difference between the RML and CLEAN images is in the CS J=7-6 emission; the disk emission stands out more prominently in the RML images than in the CLEAN images, which are comparatively noise dominated. This is due to the sparsity regularization incorporated into the RML model, which reduces the number of image pixels with non-zero values and effectively suppresses background noise in the image. 

\subsection{J1842}
The CLEAN images of J1842 show a NKF in the $^{12}$CO J=3-2 emission centered at roughly $v=5.2$ km/s. Figure \ref{fig:j1842} shows a comparison of the RML and CLEAN images of J1842 for the continuum-subtracted $^{12}$CO J=3-2, $^{13}$CO J=3-2, and CS J=7-6 data. All of the RML images in Figure \ref{fig:j1842} were made with $\lambda_{\rm{ent}} = 8\times 10^{-6}$, $\lambda_{\rm{TSV}} = 5\times 10^{-5}$, and $\lambda_{\rm{spa}} = 5\times 10^{-5}$.

The RML images are sharper than the corresponding CLEAN images, and the NKF seen in the $^{12}$CO J=3-2 CLEAN images is clearly reproduced in the RML images. This feature is consistent with a 1 $M_{\rm{Jup}}$ planet embedded in the disk \citep{Pinte_exoALMA}. Like J1615, the CS J=7-6 emission for J1842 is dominated by noise and appears more prominently in the RML images due to the sparsity regularization. However, the structure of the CS J=7-6 emission is not clear in either the RML or CLEAN images at this sensitivity; further spectral averaging could reveal the CS J=7-6 morphology with more clarity. 

\begin{figure*}
\centering
\includegraphics[width=\linewidth]{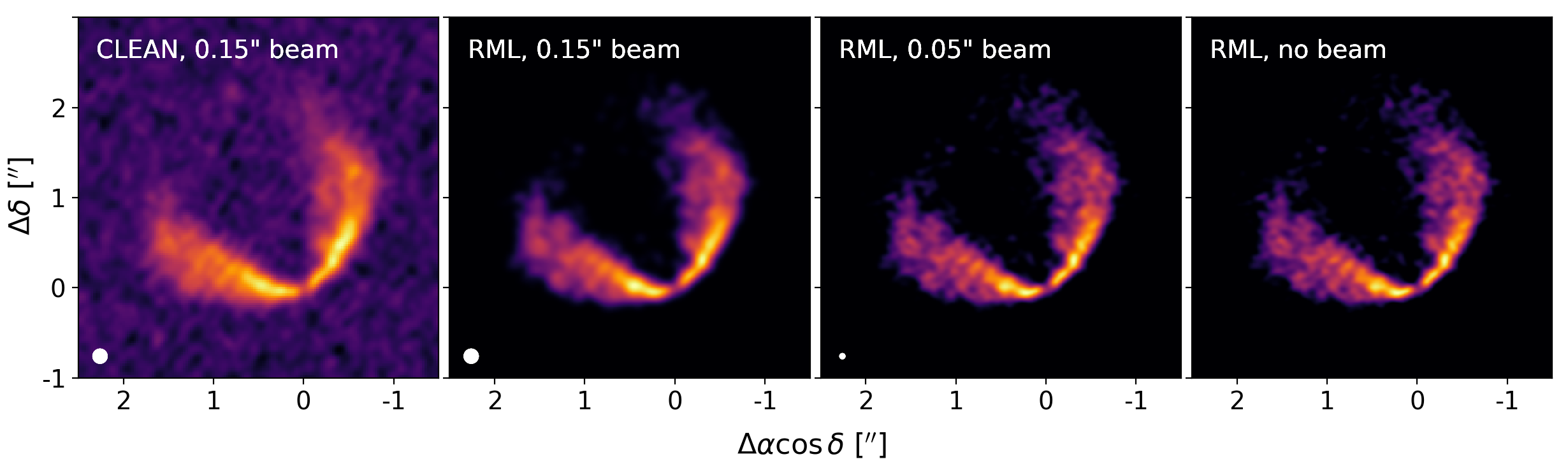}
\caption{Left to right: fiducial 0\farcs15 CLEAN, 0\farcs15 RML, 0\farcs05 RML, and native RML (no restoring beam) images. We show each image product for the continuum-subtracted $^{12}$CO J=3-2 emission in J1842 at $v=5.2$ km/s, with each panel plotted on an independent color scale. While beam convolution is not necessary in the RML imaging process, it can be useful to enable more direct comparisons between CLEAN and RML, as well as for constraining the resolution of the RML images. Beam-convolved images are also necessary for compatibility with some analysis software (e.g. tasks that require information about the restoring beam, which does not exist in the native RML image).
}
\label{fig:convolved}
\end{figure*}

\subsection{LkCa~15}

The CLEAN images of LkCa~15 show a NKF in the $^{12}$CO J=3-2 emission centered at roughly $v=6.9$ km/s. Figure \ref{fig:lkca15} shows a comparison of the RML and CLEAN images for the continuum-subtracted $^{12}$CO J=3-2, $^{13}$CO J=3-2, and CS J=7-6 data. All of the RML images in Figure \ref{fig:lkca15} were made with $\lambda_{\rm{ent}} = 0$, $\lambda_{\rm{TSV}} = 5\times 10^{-4}$, and $\lambda_{\rm{spa}} = 1\times 10^{-5}$.

The NKF in the $^{12}$CO J=3-2 emission is present in both the RML and CLEAN images, and appears across all 3 channels pictured. This feature is consistent with a 2 $M_{\rm{Jup}}$ planet embedded in the disk \citep{Pinte_exoALMA}. The RML images reproduce all features seen in the fiducial CLEAN images, though the CLEAN images are slightly sharper. The RML feature more uniform background emission due to the sparsity regularization.

\subsection{SY~Cha}

The CLEAN images of SY~Cha show a NKF in the $^{12}$CO J=3-2 emission, centered at roughly $v=3.6$ km/s. Figure \ref{fig:sycha} shows a comparison of the RML and CLEAN images of SY~Cha for the continuum-subtracted $^{12}$CO J=3-2, $^{13}$CO J=3-2, and CS J=7-6 data. All of the RML images in Figure \ref{fig:sycha} were made with $\lambda_{\rm{ent}} = 0$, $\lambda_{\rm{TSV}} = 5\times 10^{-4}$, and $\lambda_{\rm{spa}} = 1\times 10^{-5}$.

The emission in the RML images appears generally smoother and more blurred compared to the CLEAN images, but all features seen in the CLEAN images are reproduced, including the NKF in multiple channels. This feature is consistent with a 5 $M_{\rm{Jup}}$ planet embedded in the disk \citep{Pinte_exoALMA}. The sparsity regularization heavily suppresses background noise in the RML images compared to the CLEAN images, particularly in the fainter $^{13}$CO J=3-2 and CS J=7-6 emission.


\section{Discussion} \label{sec:discussion}

\begin{figure*}
\centering
\includegraphics[width=\linewidth]{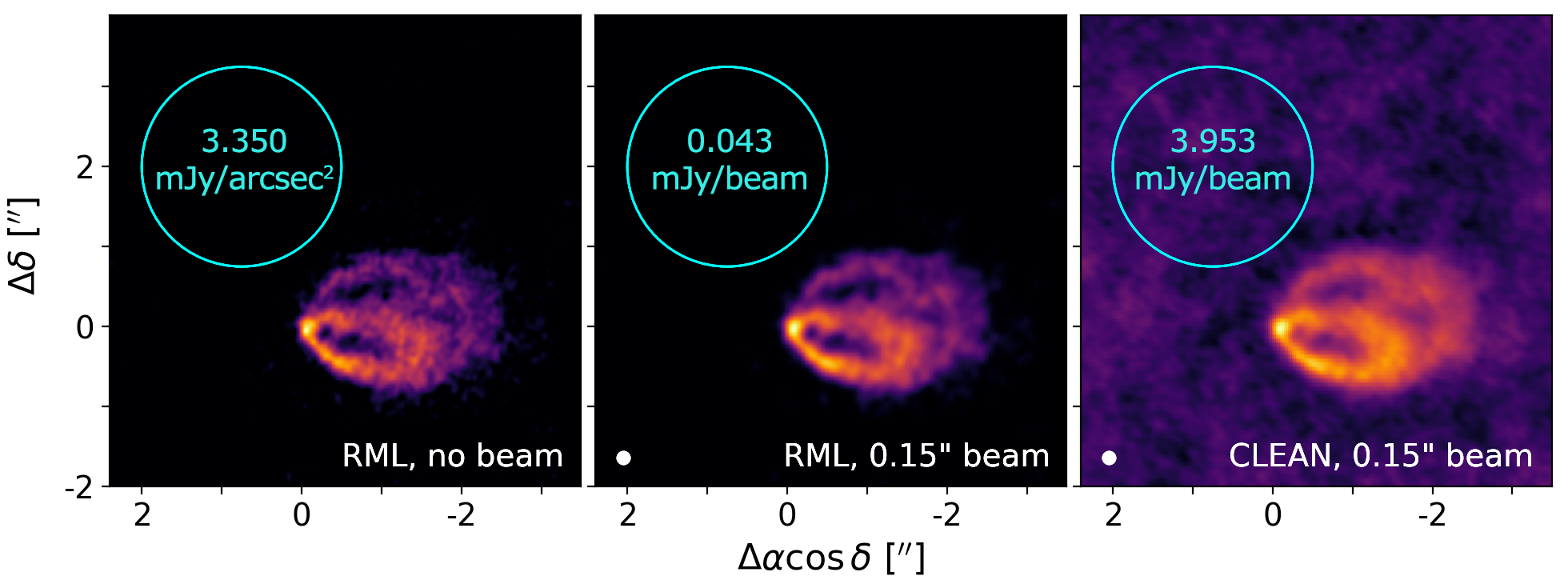}
\caption{Three different image products showing continuum-subtracted $^{12}$CO J=3-2 emission in AA Tau at $v=7.8$ km/s. We estimate the noise by taking the RMS of a signal-free region, shown with the cyan circles. Before beam convolution, the native RML images are in units of Jy/arcsec$^{2}$ (left). To enable a more direct comparison of the noise properties between RML and CLEAN images, we calculate the RMS in the RML image convolved with a circular 0\farcs15 beam (center) along with the fiducial CLEAN image which also has a circular 0\farcs15 beam (right). The RMS in the 0\farcs15 RML image is nearly two orders of magnitude lower than the RMS in the 0\farcs15 fiducial CLEAN image (0.043 mJy/beam and 3.953 mJy/beam, respectively), demonstrating that regularization can suppress the background noise far below what is expected from instrumental thermal noise.}
\label{fig:rmsnoise}
\end{figure*}

\subsection{Resolution and Noise of RML Images} \label{sec:resnoise}

Because RML images do not need to be convolved with a restoring beam (as is common practice in the final step of synthesizing CLEAN images), the spatial resolution of RML images can be difficult to define precisely. Taking the extra step to convolve RML images with a restoring beam is a straightforward way to place a firm resolution limit on the image. While RML images can often achieve a modest degree of super-resolution, the theoretical limit is roughly a factor of 4 compared to the nominal resolution of the interferometer $R_{\min }=\lambda / b_{\max}$, where $b_{\max}$ is the longest baseline in the array. However, given that there is no measured information on the scale of these superresolved features, this is only possible for data with extremely high signal-to-noise ratios and ($u,v$) coverage combined with regularizers that are well-matched to the source structure \citep[e.g.][]{Narayan_1986,Holdaway_1990, Honma_2014}. Some RML imaging applications to EHT data have achieved resolutions of 25-30\% of the diffraction limit \citep{Akiyama_2017a, Akiyama_2017b}, while others find resolutions of $\sim30$\% of the CLEAN beam \citep{Kuramochi_2018}. Studies that apply RML imaging techniques to ALMA protoplanetary disk observations typically quote the resolution in relation to the CLEAN beam (which depends on the data weighting and is often significantly larger than the diffraction limit), achieving angular resolutions of up to 1/3 the nominal CLEAN beam \citep[e.g.][]{Yamaguchi_2020, Yamaguchi_2021, Yamaguchi_2024, Zawadzki_2023}.
We thus generate two sets of beam-convolved RML cubes: one with a circular 0\farcs15 convolution kernel 
(equivalent to the fiducial CLEAN beam), and one with a circular 0\farcs05 convolution kernel 
(1/3 of fiducial CLEAN beam).

A comparison of the fiducial 0\farcs15 CLEAN, 0\farcs15 RML, and 0\farcs05 RML, and native RML (no restoring beam) images is shown in Figure \ref{fig:convolved}. We show a single channel ($v=5.2$ km/s) of $^{12}$CO J=3-2 emission for J1842 for each image product. The 0\farcs15 RML and CLEAN images show that the RML images reproduce all of the features seen in the CLEAN images, with the only major difference being the background noise suppression in the RML image. We note that a convolution kernel of a given size does not necessarily result in images with an equivalent resolution, as the true resolution of the restored image depends on both the native resolution of the RML image as well as the size of the convolution kernel. The true resolution of the convolved cube can be estimated by adding the native RML image resolution and the convolution kernel size in quadrature. For example, if the native resolution of the RML image of J1824 (rightmost panel, Figure \ref{fig:convolved}) is 0\farcs05, then convolving the image with a 0\farcs15 convolution kernel results in a restored image with a resolution of 0\farcs158. As the resolution of the native RML image cannot easily be determined, may vary spatially, and differs from source to source, we have not attempted to correct for this effect. As a result, the stated resolutions of our beam-convolved RML image cubes are lower limits on the true resolutions of the images.

With a ground truth image for comparison, it would be possible to convolve the image with a variety of beam sizes to determine the resolution which minimizes the normalized root mean square (RMS) error. This is commonly used to evaluate the quality of a reconstructed image \citep[e.g][]{Chael_2016, Akiyama_2017a, Akiyama_2017b, Kuramochi_2018, Yamaguchi_2020, Zawadzki_2023}, but without a ``true'' image it is not possible to firmly identify an optimal restoring beam size.

Estimating the noise of RML images also poses challenges, as regularization (particularly sparsity regularization) can effectively suppress background noise beyond the inherent thermal noise expected from the instrument. Figure \ref{fig:rmsnoise} shows the RMS noise in a signal-free region of three image products: a native RML, 0\farcs15 RML, and 0\farcs15 CLEAN channel of the $^{12}$CO J=3-2 emission in AA~Tau. The 0\farcs15 RML and 0\farcs15 CLEAN images offer the most direct comparison, as the image products are in the same units (Jy/beam, with identical beam sizes). The RMS noise is nearly a factor of 100 smaller in the 0\farcs15 RML image compared to the 0\farcs15 CLEAN image (0.043 mJy/beam and 3.953 mJy/beam, respectively).

This is because RML and CLEAN images are synthesized with fundamentally different procedures; while the CLEANing process results in a residual map, RML images are subject to regularizer penalties which can favor image qualities like sparsity or smoothness. With sparsity regularization in particular, when the data do not support any significant emission in a region (as in the noisy background of a high SNR image), the pixel values will tend to zero. This has the overall effect of heavily suppressing noise in signal-free regions of the image, far beyond expected values from the thermal noise of the instrument. This does not mean that the overall noise in the RML images is truly orders of magnitude better than the corresponding CLEAN products. Rather, the noise across an RML image may not be spatially uniform, as sparsity regularization only suppresses pixel values in regions where there is not likely to be any emission at all.

\subsection{Emission Surfaces and Temperatures}

We compared the non-parametric emission surfaces and radial temperature profiles, presented in full in \cite{Galloway_exoALMA}, between the CLEAN cubes and the RML cubes. To compare the two, we used RML cubes of the same spatial and spectral resolution as those used for the CLEAN cube analysis (0\farcs15, 100 m/s). We use the non-continuum-subtracted cubes as to not underestimate the temperature. 

We follow identical procedures to those outlined in Section 3 of \cite{Galloway_exoALMA} to obtain the RML emission surfaces. The procedure is summarized as follows. We utilize the open-source Python package \disksurf{}, which obtains emission surfaces via the method first demonstrated in \cite{pinte_IMLup_2018A&A...609A..47P}. To trace the emission surface, \disksurf{} locates intensity maxima along a vertical line intersecting isovelocity contours. For disks of moderate inclination, the front and back surfaces will be well-separated, allowing us to distinguish between the front and back surfaces, and to trace the $r-z$ location of emission. As in \cite{Galloway_exoALMA}, we do this process over five iterations, each time increasing the minimum signal to noise from 1.0 to 5.0 for $^{12}$CO J=3-2, and 1.0 to 3.5 for $^{13}$CO J=3-2. Minimal masking is applied to the surfaces so that we can assess the differences between CLEAN and RML. This allows us to obtain the `raw' surfaces, comprising $r-z$ points. We completed this procedure with the RML cubes of AA Tau, J1615, J1842, LkCa 15, and SY Cha for both $^{12}$CO J=3-2 and $^{13}$CO J=3-2 emission. 

\begin{figure}
\includegraphics[width=\linewidth]{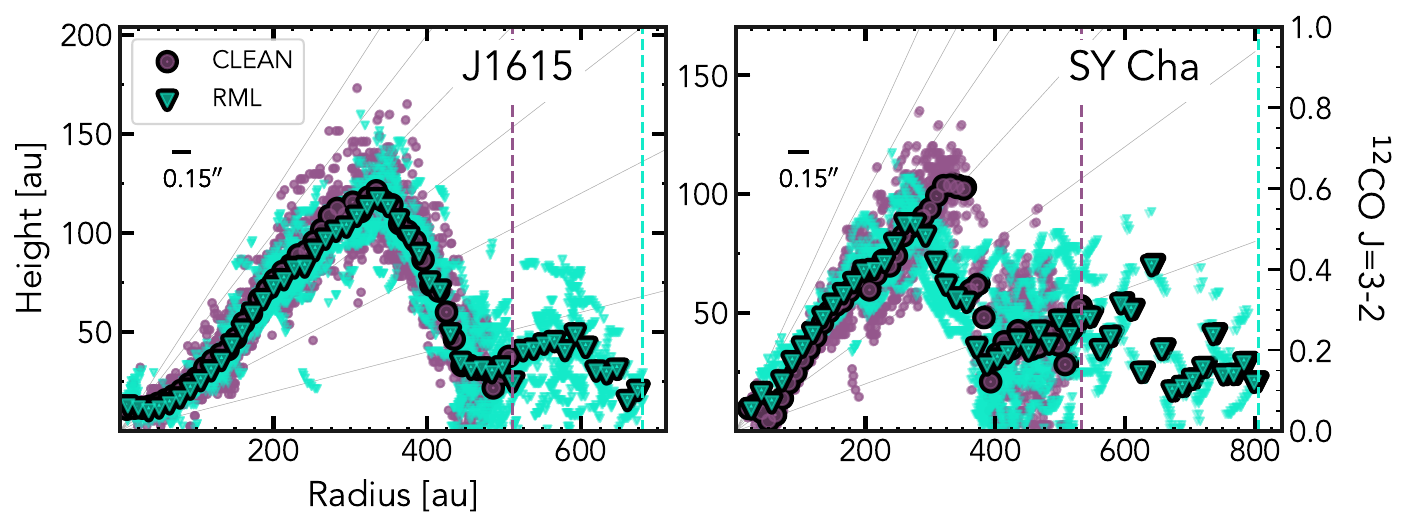}
\includegraphics[width=\linewidth]{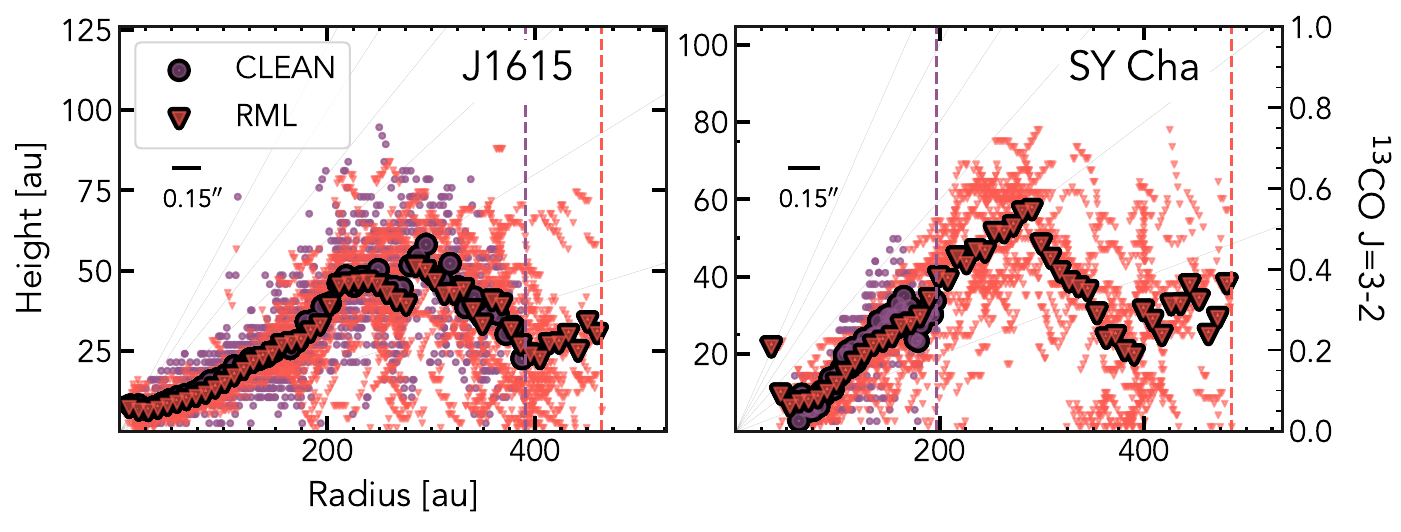}
\caption{Emission surfaces for $^{12}$CO J=3-2 and $^{13}$CO J=3-2 for J1615 and SY Cha found using the CLEAN cubes and the RML cubes. Raw $r-z$ points are shown in purple for the CLEAN data and in aqua blue and red for the RML data. The larger points show the binned surfaces. The horizontal dashed lines show the maximum surface radius, r$_{max}$. The beamsize of 0\farcs15 is shown in the upper left corner.}
\label{fig:emission_surface_RML}
\end{figure}

Figure \ref{fig:emission_surface_RML} shows the results for J1615 and SY~Cha. The top two panels show $^{12}$CO J=3-2 emission surfaces, and the bottom two show $^{13}$CO J=3-2 surfaces. The raw $r-z$ points are shown in the background in grey for the CLEAN cubes, and in aqua blue ($^{12}$CO J=3-2) and red ($^{13}$CO J=3-2) for the RML cubes. The larger points show the raw surfaces binned by a quarter of the beamsize for visual clarity. The dashed vertical lines show the maximum radial extent of the retrieved surfaces. In general, the emission surfaces are in agreement, which is also what we find for the non-pictured disks (AA~Tau, J1842, and LkCa~15). The J1615 emission surfaces both have the shame characteristic tapered power-law shape. The morphology of the SY~Cha $^{12}$CO J=3-2 RML emission surface has the biggest difference amongst the five disks tested; the surfaces begin to diverge at $\sim$300 au. Past 300 au, the CLEAN emission surface continues to rise, whereas the RML surface begins to drop down. The reasons for these morphological differences are unclear. Eventually, both surfaces drop at $\sim$400 au. 

Additionally, the RML emission surfaces tend to extend further out in both $^{12}$CO J=3-2 and $^{13}$CO J=3-2. This can be seen in nearly all of the surfaces shown in Figure \ref{fig:emission_surface_RML}. One of the most dramatic differences is in SY~Cha $^{12}$CO J=3-2 and $^{13}$CO J=3-2 emission, whose RML surface extends over 200 au further than that found using the CLEAN cube for both molecules. The effect is less pronounced in J1615, but there is still a $\sim$100 au size difference between the $^{12}$CO J=3-2 emission surfaces. This is likely due to the decreased RMS noise present in the RML cube versus the CLEAN cube. With lower noise, the \disksurf{} surface finding process will be able to identify more points that were previously cut off by the 5 SNR noise limit. 

\begin{figure}
\includegraphics[width=\linewidth]{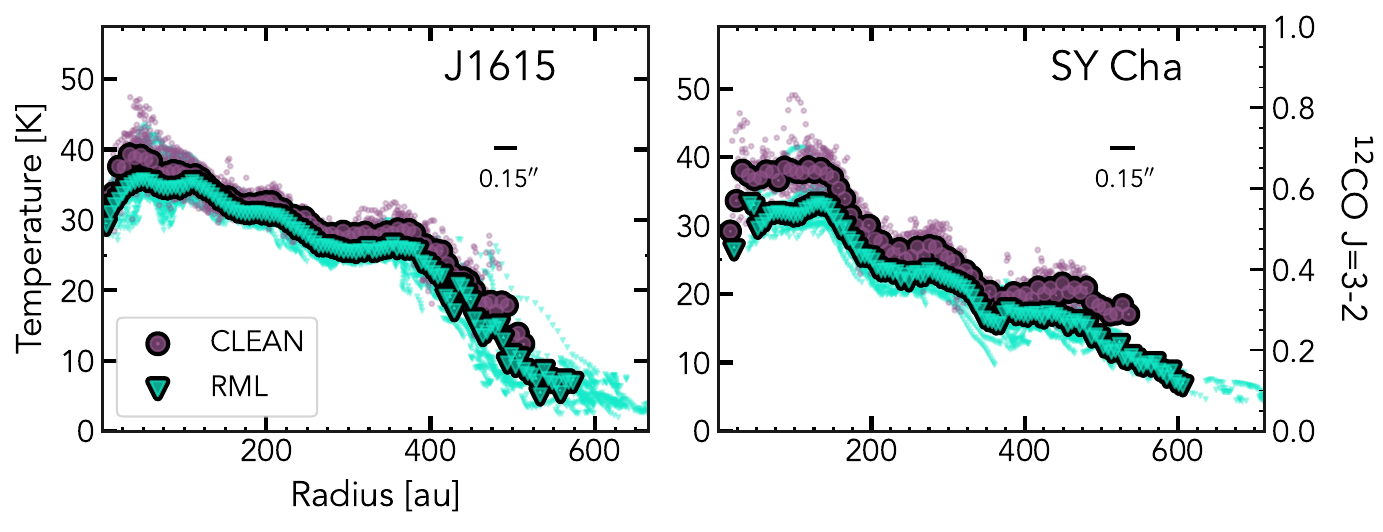}
\includegraphics[width=\linewidth]{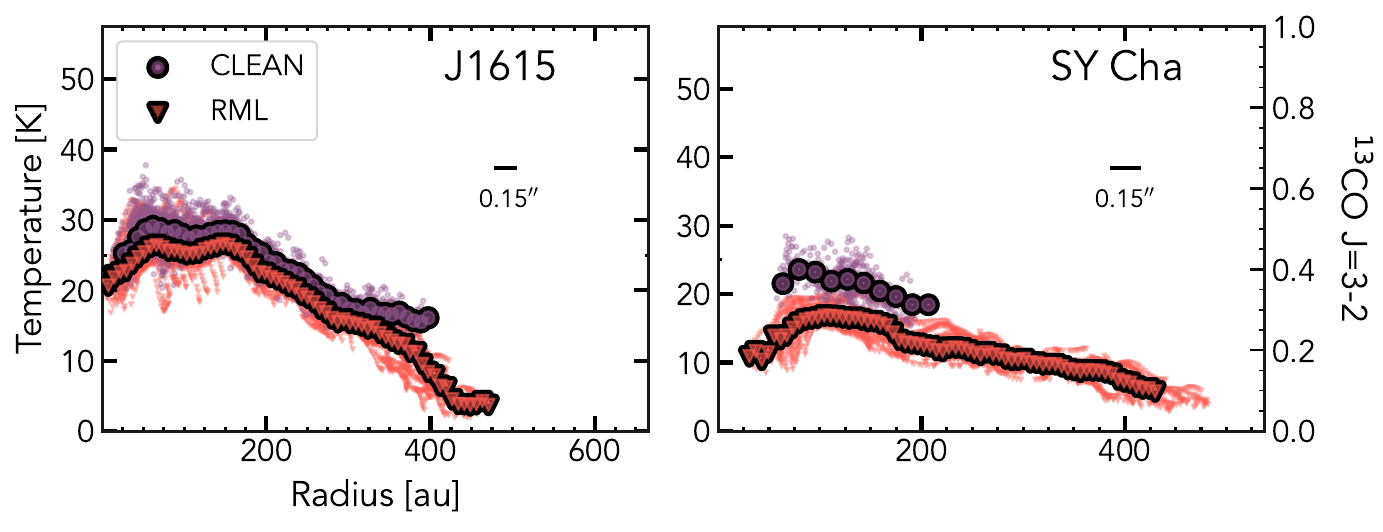}
\caption{Radial temperature profiles, found using the \disksurf{} surface points, for $^{12}$CO J=3-2 and $^{13}$CO J=3-2 for J1615 and SY Cha using the CLEAN and RML cubes. The larger points show the temperatures binned by a quarter of the beamsize. The beamsize of 0\farcs15 is shown in the upper left corner.}
\label{fig:temperatures_RML}
\end{figure}

In addition to the emission surfaces, we compare the radial temperature profiles between the CLEAN and RML cubes. We obtain the temperature as a function of radius using the $r-z$ points extracted by \disksurf{}. For further details, see Section 4 of \cite{Galloway_exoALMA}. Figure \ref{fig:temperatures_RML} shows a comparison of the brightness temperatures between the CLEAN cubes and the RML cubes for $^{12}$CO J=3-2 (top panels) and $^{13}$CO J=3-2 (bottom panels). For visual clarity, we have binned the data by a quarter of the beamsize for each corresponding cube. As noted in the previous paragraph, the maximum radial extent is larger for the RML cubes. We find that the temperatures derived from the RML cubes are systematically lower at each radius; this can be seen in the offset between the binned points. For the $^{12}$CO J=3-2 emission, the difference difference, on average, is only $\sim$3 K, but in some places it becomes as large as 13 K, such as in the inner radial regions of SY~Cha. For J1615 in $^{13}$CO J=3-2, the temperature difference, on average, is only 1.3 K. The $^{13}$CO J=3-2 emission in SY Cha exhibits the largest temperature difference between the CLEAN and RML cubes, with the average offset being 8 K. 

The causes of the temperature offset between the CLEAN and RML cubes may in part be due to resolution differences. To test this, we derived the temperatures using RML cubes of SY~Cha and AA~Tau convolved with multiple beam sizes (0\farcs5, 0\farcs15, and 0\farcs30). These sources were selected because the native RML images of SY~Cha seem to have a slightly worse resolution than the fiducial CLEAN cubes, while the native RML images of AA~Tau appear to be super-resolved. The SY~Cha temperatures extracted from the 0\farcs05 and 0\farcs15 RML images were indistinguishable, while the 0\farcs30 RML images yielded a slightly lower temperature. This behavior is observed to a larger extent with the CLEAN images, where the 0\farcs30 images result in significantly lower derived temperatures compared to the fiducial 0\farcs15 images (and also significantly lower than temperatures derived from any of the 0\farcs05, 0\farcs15, or 0\farcs30 RML images). The AA~Tau temperatures from the 0\farcs5, 0\farcs15, and 0\farcs30 images were consistent with each other, with only minor differences in the 0\farcs05 temperature profile. Specifically, the derived temperature profile for the 0\farcs05 cube was not consistently hotter than the 0\farcs15 and 0\farcs30 temperatures, but was hotter at some radii. This suggests that disk temperature profiles derived from RML images may be more robust to resolution differences than CLEAN images, though further analysis is needed to characterize this behavior in detail.

Another possible explanation for the temperature offset between the CLEAN and RML cubes is the lower RMS of the RML images. It may be that the additional emission surface points found with the RML cubes belong to colder regions of the disk. These points may be cut out by the SNR clip applied when using \disksurf{} on the CLEAN cubes, but with a lower-noise RML cube, they could remain, thus causing the offset. Another explanation could be the imaging process itself. The final intensities, and thus temperatures, are dependent on the imaging process. Fundamental differences between CLEAN and RML could lead to slight offsets in the intensities. Further exploration of this is needed. However, despite these offsets, we find consistent radial profile morphology, and most of the temperature differences are within the intrinsic temperature scatter of $\sim$5 K. 

\subsection{Limitations and Future Work}

In this work we have presented RML images of 7 exoALMA protoplanetary disks, with the primary goal of detecting protoplanets at high confidence. However, RML imaging can have other benefits beyond validating CLEAN images, namely, the potential for improved resolution in comparison to the CLEAN beam (see Section \ref{sec:resnoise}). We do not achieve a clear improvement in resolution in all the images presented here. This is likely due to the fact that we have treated all disks uniformly, i.e. tuning hyperparameters using strictly CV. This could be improved by repeating the CV procedure with different visibility partitioning, or by fine-tuning hyperparameters by hand following the CV process. A more individual RML exploration of each disk could result in a higher degree of super-resolution in the images, potentially reaching resolutions several times better the fiducial CLEAN images and enabling the potential to reveal new substructure not resolved in the CLEAN images. This detailed and source-specific procedure would require large amounts of human input and computational power, and thus is not practical for a first analysis of a large sample like the exoALMA data. However, future work should include detailed RML imaging of the full exoALMA dataset (and other protoplanetary disk data of comparable quality) with the explicit goal of obtaining super-resolved images and searching for previously unseen structure, particularly in disks for which we currently do not see evidence of embedded protoplanets. While the CLEAN images would not necessarily be able to support the detection of such features, comparison with other RML algorithms could provide a path toward confidently identifying features not seen with CLEAN imaging.

This would also require imaging the RML images with smaller pixel sizes to ensure that the model parametrization enables the highest potential degree of super-resolution. Given that the highest resolution data from the interferometer corresponds to resolutions of about 0\farcs06, the pixel size of 0\farcs0125 used in this study may not sufficiently sample the image for the best possible degree of super-resolution relative to the nominal resolution of the interferometer (but does not affect the results presented here). Pushing to a smaller pixel size would further increase the computational demands of the imaging process, and would require a new suite of hydrodynamical simulations for comparison. It would also be worthwhile to obtain RML images at the highest spectral resolution possible from the data (i.e. 28 m/s), as careful RML imaging may be able to mitigate the lower SNR without the spectral averaging, which would further increase the computational load. Nevertheless, this may be the best path forward for identifying the smallest protoplanets with the currently available data.

Additionally, future work should explore the impact of normalization schemes for different regularizers in greater depth. Here, we find that optimal hyperparameter values do not vary significantly across the cube, observing stable performance across channels even without explicitly normalizing the sparsity and TSV regularizers. However, a more detailed investigation of regularizer normalization for ALMA protoplanetary disk data, such as testing the various normalization factors presented in \citet{EHTCollab_2019}, could offer further improvements. In particular, incorporating normalization factors to account for flux variations across velocity space may enhance the image quality for fainter emission, such as the $^{13}$CO J=3-2, and CS J=7-6 data, or in the wings of the line emission, where small variations in the regularizer strength could have a stronger effect.  


\section{Conclusions} \label{sec:conclusions}

We have presented RML images for seven exoALMA targets. Our main conclusions are summarized below.
\begin{itemize}
    \item The RML methods independently and consistently reproduce the features seen in the fiducial CLEAN images, including all NKFs. The agreement between the two sets of independently synthesized image products suggests that these features are real, and strengthens findings related to the interpretation of these features \cite[such as those presented in][]{Pinte_exoALMA}. 
    \item Additionally, RML methods generally reproduce the azimuthally averaged emission surfaces and temperature profiles found with CLEAN cubes. The emission surface morphology and features are consistent, but the surfaces of the RML cubes extend further radially, sometimes by 200 au. We find that the temperature profiles measured with the RML cubes are systematically lower by $\sim$5 K, which varies radially and on a per-disk basis. This is likely due to the artificially high SNR of the RML cubes stemming from sparsity regularization, but further exploration is needed to fully characterize this effect. 
    \item Using cross-validation (CV) for hyperparameter tuning, we found that the set of hyperparameters which minimized the CV score (corresponding to a model with the highest predictive power) remained stable from channel to channel in a given cube, having only minor variations that did not significantly change the resulting image products.
    \item Furthermore, we find that optimal hyperparameters remain stable for all of the $^{12}$CO J=3-2, $^{13}$CO J=3-2, and CS J=7-6 images for a given source, suggesting that optimal hyperparameter values depend more on observational parameters like interferometer baselines and integration time rather than the specific emission morphologies.
\end{itemize}



For each target, the following data products will be publicly available:
\begin{itemize}
    \item Full $^{12}$CO J=3-2 RML cubes (both continuum-subtracted and non-continuum-subtracted)
    \item Full $^{13}$CO J=3-2 RML cubes (both continuum-subtracted and non-continuum-subtracted)
    \item Full CS J=7-6 RML cubes (both continuum-subtracted and non-continuum-subtracted)
\end{itemize}
Different versions will be released with the native RML resolution (no beam convolution), convolution with a 0\farcs05 restoring beam, and convolution with a 0\farcs15 restoring beam. We have also released sample imaging scripts, available directly on the \texttt{MPoL} GitHub\footnote{\url{https://github.com/MPoL-dev/examples}}.

\bibliography{references}{}
\bibliographystyle{aasjournal}

\begin{acknowledgments}
We thank the anonymous reviewers for their comments, which greatly improved this manuscript. Support for BZ was provided by The Brinson Foundation. Computations for this research were performed in part on the Pennsylvania State University’s Institute for Computational and Data Sciences’ Roar supercomputer. We acknowledge the U.S. National Science Foundation - Major Research Instrumentation Program, Deep Learning for Statistics, Astrophysics, Geoscience, Engineering, Meteorology and Atmospheric Science, Physical Sciences and Psychology (DL-SAGEMAPP) at the Institute for Computational and Data Sciences (ICDS) at the Pennsylvania State University. This work was performed in part on the OzSTAR national facility at Swinburne University of Technology. The OzSTAR program receives funding in part from the Astronomy National Collaborative Research Infrastructure Strategy (NCRIS) allocation provided by the Australian Government, and from the Victorian Higher Education State Investment Fund (VHESIF) provided by the Victorian Government.

This paper makes use of the following ALMA data: ADS/JAO.ALMA\#2021.1.01123.L. ALMA is a partnership of ESO (representing its member states), NSF (USA) and NINS (Japan), together with NRC (Canada), MOST and ASIAA (Taiwan), and KASI (Republic of Korea), in cooperation with the Republic of Chile. The Joint ALMA Observatory is operated by ESO, AUI/NRAO and NAOJ. The National Radio Astronomy Observatory is a facility of the National Science Foundation operated under cooperative agreement by Associated Universities, Inc. We thank the North American ALMA Science Center (NAASC) for their generous support including providing computing facilities and financial support for student attendance at workshops and publications.

JB acknowledges support from NASA XRP grant No. 80NSSC23K1312. MB, DF, JS have received funding from the European Research Council (ERC) under the European Union’s Horizon 2020 research and innovation programme (PROTOPLANETS, grant agreement No. 101002188). Computations by JS have been performed on the `Mesocentre SIGAMM' machine,hosted by Observatoire de la Cote d’Azur. PC acknowledges support by the Italian Ministero dell'Istruzione, Universit\`a e Ricerca through the grant Progetti Premiali 2012 – iALMA (CUP C52I13000140001) and by the ANID BASAL project FB210003. SF is funded by the European Union (ERC, UNVEIL, 101076613), and acknowledges financial contribution from PRIN-MUR 2022YP5ACE. MF is supported by a Grant-in-Aid from the Japan Society for the Promotion of Science (KAKENHI: No. JP22H01274). CH acknowledges support from NSF AAG grant No. 2407679. TH is supported by an Australian Government Research Training Program (RTP) Scholarship. JDI acknowledges support from an STFC Ernest Rutherford Fellowship (ST/W004119/1) and a University Academic Fellowship from the University of Leeds. A.I. acknowledges support from the National Aeronautics and Space Administration under grant No. 80NSSC18K0828. Support for AFI was provided by NASA through the NASA Hubble Fellowship grant No. HST-HF2-51532.001-A awarded by the Space Telescope Science Institute, which is operated by the Association of Universities for Research in Astronomy, Inc., for NASA, under contract NAS5-26555. CL has received funding from the European Union's Horizon 2020 research and innovation program under the Marie Sklodowska-Curie grant agreement No. 823823 (DUSTBUSTERS) and by the UK Science and Technology research Council (STFC) via the consolidated grant ST/W000997/1. CP acknowledges Australian Research Council funding  via FT170100040, DP18010423, DP220103767, and DP240103290. GR acknowledges funding from the Fondazione Cariplo, grant no. 2022-1217, and the European Research Council (ERC) under the European Union’s Horizon Europe Research \& Innovation Programme under grant agreement no. 101039651 (DiscEvol). H-WY acknowledges support from National Science and Technology Council (NSTC) in Taiwan through grant NSTC 113-2112-M-001-035- and from the Academia Sinica Career Development Award (AS-CDA-111-M03). GWF acknowledges support from the European Research Council (ERC) under the European Union Horizon 2020 research and innovation program (Grant agreement no. 815559 (MHDiscs)). GWF was granted access to the HPC resources of IDRIS under the allocation A0120402231 made by GENCI. AJW has received funding from the European Union’s Horizon 2020 research and innovation programme under the Marie Skłodowska-Curie grant agreement No 101104656. TCY acknowledges support by Grant-in-Aid for JSPS Fellows JP23KJ1008. Views and opinions expressed by ERC-funded scientists are however those of the author(s) only and do not necessarily reflect those of the European Union or the European Research Council. Neither the European Union nor the granting authority can be held responsible for them. 

\end{acknowledgments}

\software{Astropy \citep{astropy_2013, astropy_2018}, PyTorch \citep{PyTorch_2019}, MPoL \citep{mpol_2021, Zawadzki_2023, mpol_2025}, CASA \citep{McMullin_2007}, disksurf \citep{disksurf_2021JOSS....6.3827T}}

\end{document}